\documentclass[aps,prc,twocolumn,fleqn]{revtex4-1}
\usepackage[pdftex]{graphicx}
\usepackage{amsmath,amssymb}
\usepackage[colorlinks=true, pdfstartview=FitV, linkcolor=red, citecolor=blue, urlcolor=blue]{hyperref}
\usepackage{caption}
\begin{document}
\title{Longitudinal Dynamics of High Baryon Density Matter in High Energy Heavy-Ion Collisions}
\date{\today}
\author{Ming Li}
\affiliation{School of Physics and Astronomy, University of Minnesota,
Minneapolis, Minnesota 55455, USA}
\author{Chun Shen}
\affiliation{Department of Physics and Astronomy, Wayne State University, Detroit, Michigan, USA}
\affiliation{RIKEN BNL Research Center, Brookhaven National Laboratory, Upton, NY 11973, USA}
\begin{abstract}
In high energy heavy-ion collisions, the two colliding nuclei pass through each other leaving behind an almost baryon free central rapidity region. Most of the baryons are carried away by the nuclear remnants and are located in the so-called fragmentation regions. In previous papers \cite{Li:2016wzh,Li:2018ini}, it has been argued that very high baryon densities, more than ten times larger than the  normal nuclear density, can be achieved in these fragmentation regions. In this paper, we assume the high baryon density matter is thermalized at the same time as the baryon-free quark-gluon plasma in the central rapidity region. We perform a 1+1D (temporal + longitudinal) hydrodynamic simulation covering both the fragmentation regions and the central rapidity region with the baryon diffusion equation included. Baryons are found to diffuse from the fragmentation regions to the central rapidity region driven by fugacity gradients. The baryon chemical potential at freezeout monotonically increases from the central rapidity region to the fragmentation regions, suggesting a rapidity scan in high energy heavy-ion collisions might be helpful in searching for the critical point of the QCD phase diagram.  

\end{abstract}

\maketitle

\section{Introduction}
Understanding the QCD phase diagram is one of the main goals of the heavy-ion collision
experiments. For heavy-ion collisions at high collisional energy, such as the top RHIC
energy and the energies attainable at LHC, the Quark-Gluon Plasma (QGP) created in the
central region has almost zero net baryons. Lattice simulations demonstrated that at zero
baryon chemical potential the transition from the deconfined state of quarks and gluons
to the confined state of hadrons is a rapid, smooth crossover \cite{Aoki:2006we}. At
finite baryon chemical potential, the nature of the transition is unknown from the
Lattice approach due to the intrinsic sign problem, although there are some theoretical
model-dependent predictions on the nature of possible phase transitions
\cite{Stephanov:2007fk}. To create the QGP experimentally with finite baryon chemical
potential, the conventional approach is to lower the center-of-momentum collision energy
and measure observables around zero rapidity. For example, the Beam Energy Scan
porgram at RHIC carries out a series of Au+Au collisions ranging from
$\sqrt{s_{NN}} = 200\,\rm{GeV}$ down to $7.7\,\rm{GeV}$ with the corresponding baryon chemical
potential at chemical freezeout being $\mu_B \simeq 20\,\rm{Mev}$ to $ 420\,\rm{MeV}$ \cite{Adamczyk:2017iwn}.
Even larger values of the baryon chemical potential can be achieved by further lowering the
collision energy in the proposed BES phase III fixed target experiments and also in the
FAIR and the NICA experiments in Europe \cite{Ablyazimov:2017guv,Sissakian:2009zza}. This approach is based on the observation that at lower collision energies, more baryons are stopped in the central rapidity region while at higher collision energies, the central rapidity region becomes more transparent \cite{Bearden:2003hx, Arsene:2009aa}. One possible problem with creating QGP with finite baryon chemical potential using low energy heavy-ion beams is that the energy deposited in the central region might not be large enough to create a deconfined state of quarks and gluons. Instead, it might just produce a system of hadrons. In addition, modeling the initial states of low energy heavy-ion collisions are more involved than that of the high energy collisions \cite{Shen:2017bsr,Shen:2017fnn}. In high energy heavy-ion collisions, however, an alternative approach in scanning through the rapidities might be helpful in exploring the QCD phase diagram at finite baryon chemical potential. 

Depending on the reference frame in which high energy heavy-ion collisions are described, there are two different paradigms \cite{Bjorken:1982qr,Anishetty:1980zp}. Bjorken's paradigm \cite{Bjorken:1982qr}, which is presented in the center-of-momentum frame, concerns the boost-invariant central rapidity region and argues that the energy density deposited is large enough for the formation of the QGP. This paradigm completely ignores the baryons which are carried away from the central rapidity region by the nuclear remnants. Also, the colliding nuclei are highly Lorentz contracted in the center-of-momentum Cartesian reference frame. Attempting to resolve the longitudinal structure of the nuclear remnants turns out to be unrealistic in the Cartesian reference frame. The AKM paradigm (Anishetty, Koehler and McLerran) \cite{Anishetty:1980zp}, which is characterized in the lab frame of a fixed-target experiment, concerns the fragmentation regions and argues that the baryon densities inside the target fireball after the collisions are enhanced due to the nuclear compression. The energy densities of the target fireball could also be large enough for the formation of the QGP. In the lab frame of a fixed-target experiment, before the collision, the projectile nucleus is highly Lorentz contracted while the target nucleus retains its longitudinal structure. These two different paradigms describe the same physical process and complement each other. Physics that are Lorentz invariant should be incorporated in either paradigm to be considered complete. In particular, the nuclear compression and its resulting enhancement of baryon densities emphasized in the AKM paradigm, which were later verified in \cite{Gyulassy:1986fk}, should be included in the fragmentation regions of Bjorken's paradigm. In references \cite{Li:2016wzh,Li:2018ini}, the initial baryon densities and energy densities achievable in the fragmentation regions of central Au+Au collisions at $\sqrt{s_{NN}} = 200\,\rm{GeV}$ were estimated using the McLerran-Venugopalan model \cite{McLerran:1993ni, McLerran:1993ka}. The baryon densities were found to be more than ten times larger than the normal nuclear density while the energy densities are also more than ten times larger than the critical energy density for the formation of the QGP. With such large initial baryon and energy densities, quarks and gluons are very likely to be liberated and thermalized in the fragmentation regions just like that happened in the central rapidity region. The exact mechanism of the thermalization in the fragmentation regions is expected to be as challenging as the early thermalization problem in the central rapidity region \cite{Baier:2000sb,Kurkela:2015qoa} and beyond the scope of the current paper. 

Assuming the high baryon densities in the fragmentation regions of high energy heavy-ion collisions are thermalized, we focus on the subsequent dynamical evolution of the high baryon density matter. Unlike the QGP with almost zero baryon chemical potential in the central rapidity region, the high baryon density matter in the fragmentation regions naturally involves finite/large baryon chemical potential. Probing the fragmentation regions of high energy heavy-ion collisions thus provides an alternative experimental approach in studying quark-gluon plasma with finite/large baryon chemical potential. In this sense, exploring the dynamical evolution of the high baryon density matter is important as it will reveal the redistribution of baryons as the system evolves. In this paper, we will concentrate on the central collision of Au+Au at $\sqrt{s_{NN}} = 200\,\rm{GeV}$. To be precise, we use the 1+1D (temporal + longitudinal) hydrodynamic model to simulate the subsequent space-time evolution of the high baryon density matter. In doing so, we combine the QGP in the central rapidity region and the high baryon density QGP at the fragmentation region as one fluid and examine the longitudinal dynamics of the baryon diffusion.

The paper is organized as follows. In Sect. \ref{sec:diffsive_equations}, the relativistic diffusive equations to simulate the high baryon density matter are presented. Section \ref{sec:input_to_hydrodynamics} discusses the input to the hydrodynamic equations including the initial conditions, the equation of state, the transport coefficients and the final freezeout. Results are given and discussed in Sect. \ref{sec:longitudinal_dynamics}. 

\section{Relativistic Diffusive Hydrodynamic Equations}
\label{sec:diffsive_equations}

We briefly summarize the relativistic hydrodynamic equations and notations used in the paper. The readers are refered to  \cite{Romatschke:2009im,Gale:2013da,Jeon:2015dfa,Jaiswal:2016hex,Romatschke:2017ejr,Denicol:2012cn} for more thorough discussions on the application of hydrodynamics in relativistic heavy-ion collisions. The hydrodynamic equations are
\begin{equation}\label{eq:hydro_baryon_conservation}
\partial_{\mu} T^{\mu\nu} =0,\quad \partial_{\mu} J^{\mu}_B =0 .
\end{equation}
with the energy-momentum tensor $T^{\mu\nu}$ and the baryon current for an relativistic diffusive fluid having the expressions
\begin{equation}\label{eq:hydro_EM}
\begin{split}
&T^{\mu\nu} = \varepsilon u^{\mu}u^{\nu} - P \Delta^{\mu\nu},\\ 
&J^{\mu}_{B} = n_B u^{\mu} + V^{\mu}.\\
\end{split}
\end{equation}
Here $\varepsilon\equiv \varepsilon(x)$, $P\equiv P(x)$, $u^{\mu}\equiv u^{\mu}(x)$,
$n_B\equiv n_B(x)$ and $V^{\mu}\equiv V^{\mu}(x)$ are the local energy density, the
thermal pressure, the velocity field, the local baryon density and the diffusive baryon
current, respectively. The time-like velocity field $u^{\mu}$ is chosen to be along the
energy current satisfying $T^{\mu}_{\,\,\nu}u^{\nu} = \varepsilon u^{\mu}$, the so-called
Landau frame. The velocity field is further normalized to be $u^{\mu}u_{\mu}=1$. Raising
or lowering tensor indexes are through the Minkowski metric $g_{\mu\nu} =
\mathrm{diag}(1,-1,-1,-1)$. The $\Delta^{\mu\nu} = g^{\mu\nu} - u^{\mu}u^{\nu}$ is a
symmetric projection operator orthogonal to the velocity field
$\Delta^{\mu\nu}u_{\nu}=\Delta^{\mu\nu}u_{\mu}=0$. The diffusive part of the baryon
current is perpendicular to the four velocity $u_{\mu}V^{\mu}=0$. Also, $V^{\mu}(x)$
is the diffusive correction to the baryon current of ideal fluid $J^{\mu}_{B,
\mathrm{id}} = n_B u^{\mu}$. It allows baryon charges to flow with a different velocity
compared to the local energy density. The baryon diffusive current $V^{\mu}(x)$ follow its own dynamical equation that asymptotically approaches the Navier-Stokes limit \cite{Jeon:2015dfa}
\begin{equation}\label{eq:viscous_diffusive_equations}
\tau_V\Delta^{\mu\nu} DV_{\nu} + V^{\mu} = \kappa_B I^{\mu} + \mathcal{X}^{\mu}\,.
\end{equation}
Here $\tau_V$ is the relaxation time for the baryon diffusive current and $\kappa_B$ is the baryon transport coefficient.  The comoving time derivative $D=u^{\mu}\partial_{\mu}$ reduces to $\partial_t$ in the fluid local rest frame where $u^{\mu}=(1,0,0,0)$. The  $\nabla^{\mu}=\Delta^{\mu\nu}\partial_{\nu}$ is the space-like gradient.  The vector $I^{\mu} = \nabla^{\mu}(\mu_B/T)$ with $\mu_B$ and $T$ the baryon chemical potential and temperature is due to the gradient of the fugacity. The $\mathcal{X}^{\mu}$ represents other possible terms that can be higher oder gradients of hydrodynamical and thermodynamical variables. For example, Israel and Stewart \cite{Israel:1979wp} predicted there are terms  $-\tau_V\theta V^{\mu}$ in $\mathcal{X}^{\mu}$. How many terms there are and what the relative importance of these terms to the study of QGP are still under active research and debate \cite{Denicol:2012cn, Baier:2007ix}.\\

The study of the QGP in the central region using hydrodynamics usually ignores the baryon current conservation equation in Eq. \eqref{eq:hydro_baryon_conservation} and the associated baryon diffusion equation Eq. \eqref{eq:viscous_diffusive_equations} because the net baryon charge is close to zero and boost-invariance is usually assumed in high energy heavy-ion collisions. Most of the efforts are focused on the effects of shear viscosity in reproducing the anisotropic flow data and extracting the temperatue dependence of shear viscosity over entropy density \cite{Romatschke:2009im, Song:2012ua}. With the high baryon density initial conditions in the fragmentation regions of high energy heavy-ion collisions, baryon conservation and diffusion equations must be included in the hydrodynamic description of the QGP evolution, and the boost-invariance assumption becomes invalid. A full 3+1D numerical computation of Eqs.\eqref{eq:hydro_baryon_conservation} and \eqref{eq:viscous_diffusive_equations} with high baryon density initial conditions are challenging \cite{Denicol:2018wdp}. Instead, we focus on the 1+1D (temporal+longitudinal) situation and study the longitudinal dynamics of the high baryon density matter, given the initial conditions. We also ignore the viscous effects and focus on the baryon diffusion effect. This is complementary to the conventional hydrodynamic simulations of the QGP evolution in high energy heavy-ion collision which usually ignore the longitudinal dynamics by assuming boost-invariance and not taking into account the high baryon densities in the fragmentation regions. We hope this 1+1D study will give us a physical picture of the longitudinal dynamics of the baryons before a full 3+1D description is developed.

There are several studies of the 1+1D hydrodynamics for high energy heavy-ion collisions without assumping boost-invariance \cite{Satarov:2006iw, Bozek:2007qt, Monnai:2012jc, Florkowski:2016kjj}. However, none of these studies take into account the high baryon density initial conditions in the fragmentation regions.  For the baryon diffusion equation Eq.\eqref{eq:viscous_diffusive_equations}, only the fugacity gradient term is kept on the right hand side ignoring the $\mathcal{X}^{\mu}$ terms. This simplified baryon diffusion equation is sometimes known as the Cattaneo equation \cite{Cattaneo:1958} and it has been used in the study of hydrodynamic fluctuations in relativistic heavy-ion collisions \cite{Kapusta:2017hfi,Kapusta:2014dja}. We rewrite the equations using Milne coordinates $(\tau, \eta)$ defined from the Cartesian coordinates by $\tau=\sqrt{t^2-z^2}$ and $\eta = \frac{1}{2}\ln\left(\frac{t+z}{t-z}\right)$. The set of equations we are solving in terms of the Milne coordinates are
\begin{equation}\label{eq:1d_explicit}
\begin{split}
&\partial_{\tau}(\tau T^{\tau\tau}) + \partial_{\eta} \tilde{T}^{\tau\eta} + \tilde{T}^{\eta\eta} =0,\\
&\partial_{\tau}(\tau\tilde{T}^{\tau\eta}) +\partial_{\eta} \tilde{T}^{\eta\eta} + \tilde{T}^{\tau\eta} =0,\\
&\partial_{\tau}(\tau J^{\tau})+\partial_{\eta} (\tau J^{\eta}) = \mathcal{S}_B ,\\
&\partial_{\tau}(\tilde{V}^{\eta}) + \partial_{\eta}\left(\frac{u^{\eta}}{u^{\tau}} \tilde{V}^{\eta}\right) = \mathcal{S}_V.\\
\end{split}
\end{equation}
with the source terms being 
\begin{equation}\label{eq:physical_sources}
\begin{split}
&\mathcal{S}_B = -\partial_{\tau}\left(\frac{\tau^2 u^{\eta}}{u^{\tau}}\tilde{V}^{\eta}\right) - \partial_{\eta}\tilde{V}^{\eta},\\
&\mathcal{S}_V = \frac{\tilde{V}^{\eta}}{u^{\tau}}(\partial_{\tau}u^{\tau} + \partial_{\eta}u^{\eta})- \frac{1}{\tau_V}\frac{\tilde{V}^{\eta}}{u^{\tau}}\\
&\qquad-\frac{\kappa_B}{ \tau_V}[u^{\tau}\tau^{-1}\partial_{\eta}+\tau u^{\eta}\partial_{\tau}]\left(\frac{\mu_B}{T}\right).\\
\end{split}
\end{equation}
Here 
\begin{equation}
\begin{split}
&T^{\tau\tau} = (\varepsilon + P) u^{\tau}u^{\tau} - P\, ,\\
&\tilde{T}^{\tau\eta} = \tau T^{\tau\eta} = (\varepsilon + P) \tau u^{\tau}u^{\eta}\, ,\\
&\tilde{T}^{\eta\eta} = \tau^2 T^{\eta\eta} = (\varepsilon + P)\tau^2 u^{\eta}u^{\eta} + P\, ,\\
&J^{\tau} = n_B u^{\tau} ,\\
&J^{\eta} = n_B u^{\eta},\\
&\tilde{V}^{\eta} = \tau V^{\eta}. \\
\end{split}
\end{equation}
The 1+1D hydrodynamic equations Eq.\eqref{eq:1d_explicit} are written in a form that is  suitable for numerical calculations using the Central-Upwind scheme (Kurganov-Tadmor scheme) \cite{Kurganov:2016, KurganovLin:2007, KurganovTadmor:2000}. The Kurganov-Tadmor scheme is a finite-volume method in solving hyperbolic partial differential equations and is used in the MUSIC package \cite{Schenke:2010nt,Schenke:2010rr, Paquet:2015lta} for a 3+1D simulation of heavy-ion collisions.  

\section{Inputs to 1+1D Diffusive Hydrodynamic Equations}
\label{sec:input_to_hydrodynamics}

\subsection{Initial Conditions}
We assume the hydrodynamic evolution of the high baryon density matter starts from $\tau_0 = 0.6\,\rm{fm/c}$ for central Au+Au collisions at $\sqrt{s_{NN}} = 200\,\rm{GeV}$. This is the typical starting time chosen to simulate the almost baryon-free central region of high energy heavy-ion collisions using hydrodynamics \cite{Song:2010aq}. To initialize the 1+1D diffusive hydrodynamic equations Eqs.\eqref{eq:1d_explicit}, one needs to specify the energy density $\varepsilon(\eta)$, the baryon density $n_B(\eta)$, the velocity field $u^{\tau}(\eta)$ and the baryon diffusion current $V^{\eta}(\eta)$ as functions of the pseudorapidity $\eta$ at the initial time $\tau_0$. 

The high baryon density initial conditions developed in \cite{Li:2016wzh,Li:2018ini} concern the forward/backward large rapidity regions while assuming the central region around $\eta\sim 0$ to be boost-invariant. Lorentz scalars like the energy density and baryon density are independent of $\eta$ in the central region. For central Au+Au collisions at $\sqrt{s_{NN}} = 200\,\rm{GeV}$ the magnitude of the energy density is chosen to be $\varepsilon(\eta=0) = 30\,\rm{GeV/fm}^3$ in the central region \cite{Song:2010aq} while the net baryon density is chosen to be $n_B(\eta=0) = 0$. In the forward/backward rapidity regions where the high baryon densities are located, the boost-invariance assumption is invalid and the pseudorapidity dependence of $\varepsilon(\eta)$ and $n_B(\eta)$ are given by the high baryon density initial conditions in \cite{Li:2016wzh,Li:2018ini}. We have to combine the forward/backward rapidity regions with the central region to obtain the initial distributions across the whole range of the pseudorapidity. In principle, a detailed dynamical study of the interplay between the forward/backward rapidity regions and the central regions is needed to give a realistic initial distributions. Lacking such a study, we instead take a more practical approach by smoothly connecting the forward/backward rapidity regions to the central regions using a half-Gaussian distribution. The center and the width of the half-Gaussian distribution are free parameters chosen appropriately in the numerical calculations. The half-Gaussian distribution has the expression \cite{Schenke:2010nt,Hirano:2001eu,Hirano:2002ds}
\begin{equation}
H(\eta) = \mathrm{exp}\left[-\frac{(|\eta| - \eta_f/2)^2}{2\sigma^2_{\eta}}\theta(|\eta| - \eta_f/2)\right].
\end{equation}
Here $\eta_f$ characterizes the range of the central boost-invariant region and $\sigma_{\eta}$ is the Gaussian width at the boundaries $\pm\eta_f/2$ between the central region and the forward/backward rapidity regions. The $\theta(x)$ is the Heaviside step function. 

\begin{figure}[t]
	\centering
	\includegraphics[scale=0.7]{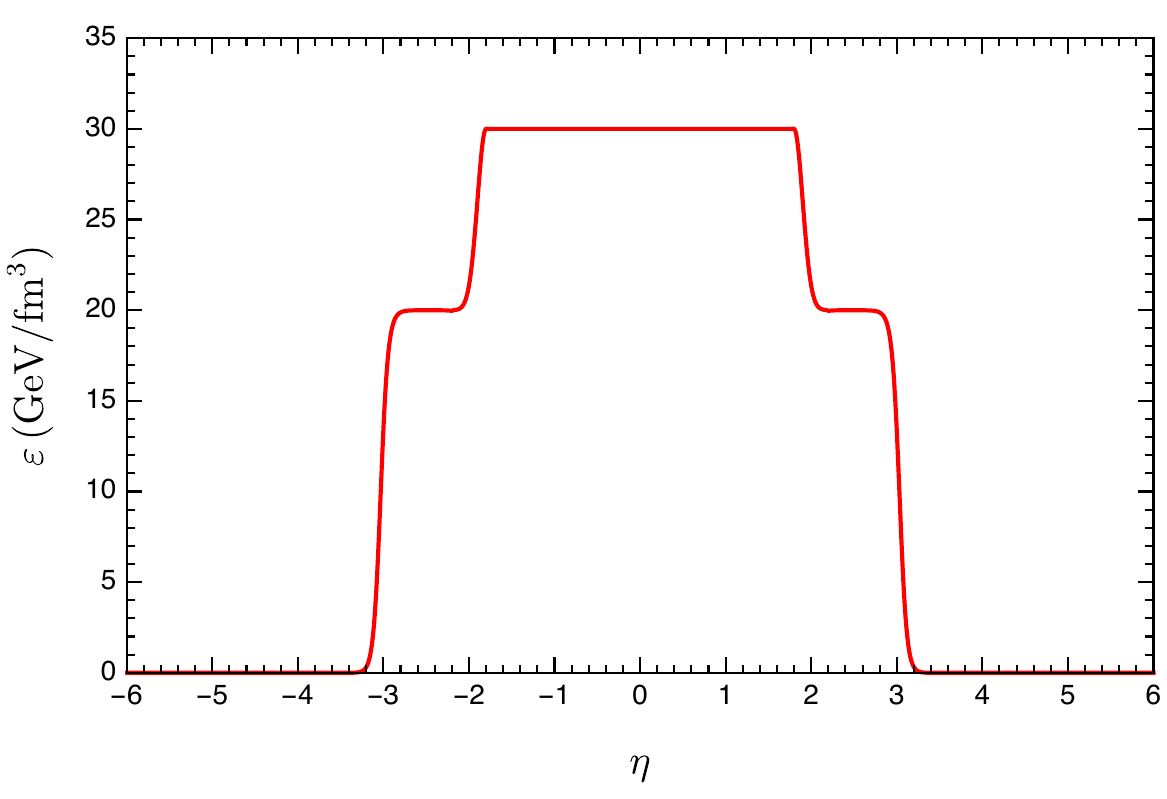}
	\caption{1+1D initial energy density distribution $\varepsilon(\eta)$ at $\tau_0=0.6\,\rm{fm/c}$.}
	\label{fig:initial_energy_distribution}
\end{figure}

\begin{figure}[t]
	\centering
	\includegraphics[scale=0.7]{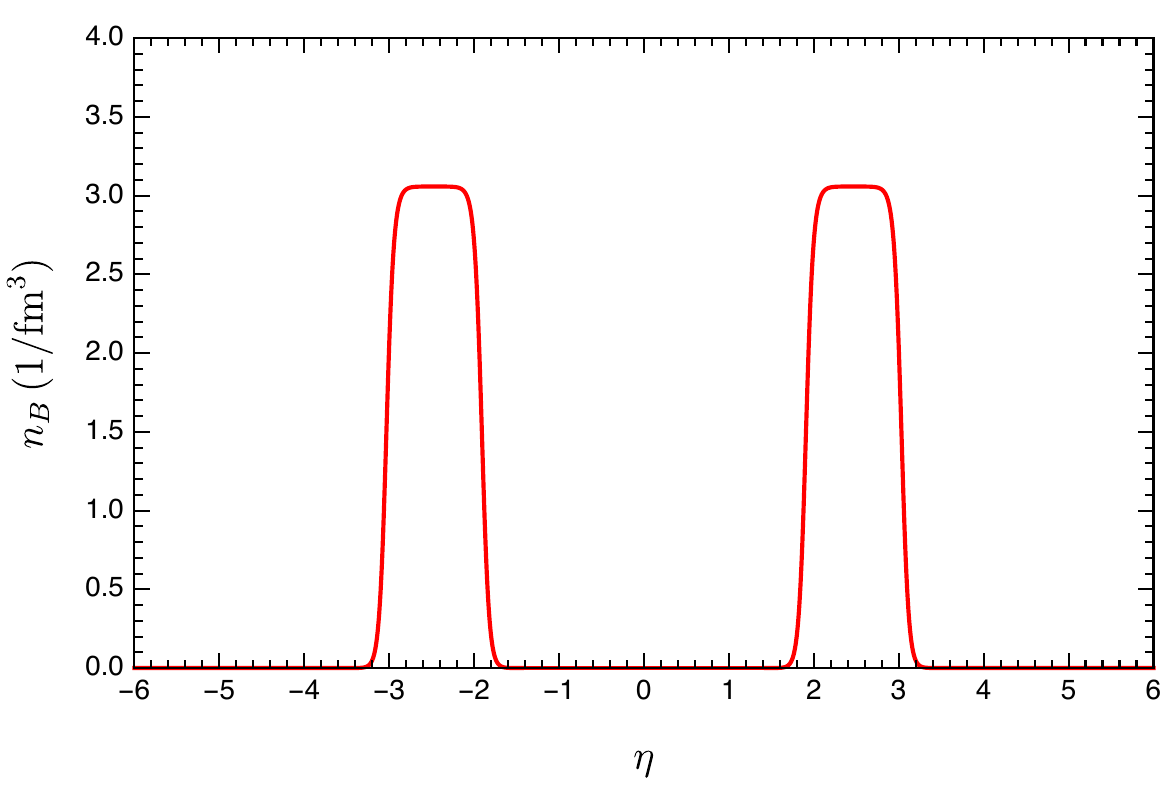}
	\caption{1+1D initial baryon density distribution $n_B(\eta)$ at $\tau_0=0.6\,\rm{fm/c}$.}
	\label{fig:initial_baryon_density_distribution}
\end{figure}

For 1+1D studies of central Au+Au collisions at $\sqrt{s_{NN}} = 200\,\rm{GeV}$, we focus on the position $r_{\perp}=0$ in the transverse plane. The initial energy density distribution $\varepsilon(\eta)$ at $\tau_0=0.6\,\rm{fm/c}$ is shown in Fig. \ref{fig:initial_energy_distribution}. The range of the central boost-invarant region is chosen to be $\eta_{f} = 3.6$ and the Gaussian width is chosen to be $\sigma_{\eta} = 0.1$.  In the central region, the energy density is $30\,\rm{GeV/fm}^3$ and smoothly connects to the forward/backward rapidity regions where the largest energy density is around $20\,\rm{GeV/fm}^3$. The rapidity regions $1.8\leq|\eta|\leq3.2$ are where the high baryon densities are located. Figure \ref{fig:initial_baryon_density_distribution} shows the initial baryon density distribution. The half-Gaussian connection has not been used for the baryon density distribution as the baryon density in the central region is also assumed to be zero in the discussions in Refs. \cite{Li:2016wzh,Li:2018ini}. For the momentum space rapidity $y(\eta)$ which characterizes the velocity fields, one might as well expect the half-Gaussian connection should apply:  $y=\eta$ for $|\eta|\leq 1.8$ and $y=2.47$ for $1.8\leq|\eta|\leq3.2$. Thermal motions of baryons inside the baryonic fireballs are ignored at the initial time and this constant rapidity $y=2.47$ reflects the overall velocities of the receding baryonic fireballs. However, thermodynamic variables are very sensitive to the change of the velocity fields; the half-Gaussian connection for $y(\eta)$ introduces artifical sudden changes of the velocity fields in the connection region. This consequently causes unphysical drops in the energy density profile as the system evolves. On the other hand, the matter in the central regions is pushed to the large rapidity regions as the system evolves and the rapidity profile quickly adjusts itself to be close to the boost-invariant distribution for the velocity fields. In this sense, we will choose the initial rapidity profile to be $y=\eta$, which is essentially equivalent to turning off initial flows and let the system adjust itself. 

\subsection{Equation of State (EoS) and Transport Coefficient}
To study the space-time evolution of the matter created both in the central region and the forward/backward rapidity regions at the same time in high energy heavy-ion collisions, the EoS should include the crossover feature at zero baryon chemical potential as well as the possible critical point and first order phase transition line at finite baryon chemical potential. Unfortunately, such an EoS has not been obtained from lattice claculations, although there are some model-dependent theoretical parameterizations \cite{Parotto:2018pwx}. In the numerical study of the hydrodynamic evolution of the high baryon density matter, we wil instead use the crossover EoS developed by Albright, Kapusta and Young in \cite{Albright:2014gva, Albright:2015uua}.  The crossover EoS is valid for finite baryon chemical potential and does not contain any critical points or first order phase tansitions. Although it might not be a realistic EoS due to the lack of the critical point and the first order phase transition line, it is smooth and analytic at finite baryon chemical potential and is sufficient for our study of high baryon density matter as long as we are not focused on any critical phenomena. 

Apart from the equation of state, we also need expressions for the baryon transport coefficient $\kappa_B$ and the relaxation time $\tau_V$ at finite baryon chemical potential. Like the equation of state, $\kappa_B$ and $\tau_V$ are determined by the microscopic dynamics of the high baryon density matter and are calculable should we knew the microscopic degrees of freedom and their interactions. However, first principle calculations of $\kappa_B$ and $\tau_V$ at finite baryon chemical potential are very difficult if near impossible. There are expressions for $\kappa_B$ and $\tau_V$ at finite baryon chemical potential from weakly coupled theory using the kinetic theory approach \cite{Albright:2015fpa, Jaiswal:2015mxa,Greif:2017byw, Denicol:2018wdp} and strongly coupled theory using the AdS/CFT correspondance \cite{Son:2006em, Rougemont:2015ona}. In \cite{Denicol:2018wdp}, the Boltzmann equation for the single particle distribution function is solved in the relaxation time approximation using the Chapman-Enskog expansion. In the massless limit using the classical Boltzmann distribution for equilibrium distribution, the baryon transport coefficient has the form 
\begin{equation}\label{eq:kappaB_kinetic_theory}
\kappa_B = \tau_V n_B\left( \frac{1}{3}\coth{\left(\frac{\mu_B}{T}\right)} - \frac{n_B T}{\varepsilon+P}\right).
\end{equation}
The relaxation time $\tau_V$ is assumed to be inversely proportional to the temperature $\tau_V = C_B/T$ with a free parameter $C_B$. It is worth emphasizing that the baryon transport coefficient Eq.\eqref{eq:kappaB_kinetic_theory} comes from the kinetic theory approach. The kinetic theory approach serves as a good description of the hadronic phase particularly in a hadronic quasiparticle model \cite{Albright:2015fpa}. This approach might be still applicable in the extremely large temperature and/or baryon chemical potential regions where the quark-gluon plasma can be approximated as a weakly-coupled system so that perturbative QCD approach is justified \cite{Jaiswal:2015mxa,Danielewicz:1984ww,Hosoya:1983xm,Gavin:1985ph,Heiselberg:1993cr}. However, one should not expect this approach to be applicable in the strongly coupled QGP regime. There are no known results of the baryon transport coefficient from the Lattice QCD calculations. Our current understanding of the baryon transport coefficient in the QGP phase across a wide range of temperatures and baryon chemical potentials is very little \cite{Floerchinger:2015efa}. In Ref. \cite{Son:2006em} the thermal conductivity of an R-charged $\mathcal{N}=4$ supersymmetric Yang-Mills theory was calculated using the AdS/CFT correspondence.  Translating into the baryon transport coefficient, it has the form
\begin{equation}\label{eq:kappaB_holographic}
\kappa_B = 2\pi \frac{Ts}{\mu_B^2}\left(\frac{n_B T}{\varepsilon + P}\right)^2. 
\end{equation}
The expression is quoted here and will be used in the numerical simulation of the 1+1D diffusive hydrodynamics merely to compare with results obtained when using Eq.\eqref{eq:kappaB_kinetic_theory}. 

\subsection{Freeze-Out}
At late times of the hydrodynamic evolution of the QGP, the temperature decreases and the system becomes dilute. The deconfined quarks and gluons recombine into confined states of hadrons. The hadronic system expands futher until the mean free path of the constituent particles becomes comparable to the typical macroscopic length size, and hydrodynamics breaks down. It would be more appropriate to describe the subsequent evolution of the hadronic system using kinetic theory, such as the hadronic cascade model UrQMD \cite{Bass:1998ca, Bleicher:1999xi}. As the system continues expanding, the chemical freeze-out and kinetic freeze-out follow when inelastic collisions, particle decays and elastic collisions gradually stop. The resulting stable particles freely travel to the detectors. While it is more realistic to use a hybrid model (hydrodynamics + transport model) to describe the particle productions, in this simple 1+1D study, we use the Cooper-Frye freeze-out mechanism \cite{Cooper:1974mv} to estimate the final particle distribution which comes directly from the end of the hydrodynamics. We also take into account of the resonance decay contribution to the particle production.

\subsubsection{Direct Thermal Production}
At the moment that hydrodynamics is assumed to break down, particles are thermally emitted from the individual fluid cells.  The momentum space distributions of the emitted particles are given by the Cooper-Frye formula \cite{Cooper:1974mv}

\begin{equation}\label{eq:cooper_frye}
E\frac{dN^{th}_i}{d^3\mathbf{p}} = \frac{dN^{th}_i}{dy d^2\mathbf{p}_{T}} =\frac{ g_i}{(2\pi)^3} \int_{\Sigma} f^i(x,p)\, p^{\mu}d^3\Sigma_{\mu}.
\end{equation}
Here $\Sigma$ represents the freeze-out hypersurface in the four dimensional spacetime. The $d^3\Sigma_{\mu}$ is the normal four vector of the hypersurface. Its magnitude tells the size of the infinitesimal hypersurface patch while the four vector tells the direction of the normal vector associated with each of these hypersurface patches. For the particle species labeled by $i$, the $g_i$ is the degeneracy factor and $f^i(x,p)$ is the phase space distribution which can be decomposed into the equilibrium part and the nonequilibrium correction 
\begin{equation}\label{eq:equil_nonequil_distribution}
f^i(x,p) = f^i_0(x,p) + \delta f^i(x,p)\, .
\end{equation}
The local equilibrium distribution $f^i_0(x,p)$ is
\begin{equation}
f^i_0(x,p) = \frac{1}{e^{(u^{\mu}p_{\mu} -b_i\mu_B)/T}\pm 1}
\end{equation}
Here $b_i$ represents the number of baryon charges that the particle species $i$ carries. The $u^{\mu}(x)$, $\mu_B(x)$ and $T(x)$ are macroscopic functions of spacetime representing flow velocity, baryon chemical potential and temperature pertaining to the local fluid cell at $x$. The nonequilibrium correction to the distribution function $\delta f^i(x,p)$ depends on specific models used to describe the nonequilibrium processes. We only consider the baryon diffusion process. Within a kinetic theory approach in the relaxation approximation, it has the form \cite{Denicol:2018wdp}
\begin{equation}
\delta f^i(x,p) = f^i_0(x,p)(1\pm f^i_0(x,p))\left(\frac{n_B}{\varepsilon + P} - \frac{b_i}{u^{\mu}p_{\mu}}\right)\frac{p^{\mu}V_{\mu}}{\kappa_B/\tau_V}
\end{equation}

\subsubsection{Resonance Decay Production}
The Cooper-Frye mechanism not only gives the direct thermal distribution of stable hadrons such as proton and pions, it also predicts the distribution of unstable resonance particles such as the nucleon resonances $N^{\ast}$ and the $\Delta$ baryons. Ultimately, we will be interested in the final proton spectrum which partly comes from direct thermal production and partly comes from the decays of resonance baryons. For example, the decay channel $\Delta^+(1232)\rightarrow p + \pi^0$ contributes to the proton spectrum. Consider the decay process that a resonance particle labeled by $R$ decays to the particle of interest labeled by $i$ and other unidentified N particles $R\longrightarrow i+1+2+\cdots N$. 
The Lorentz-invariant momentum spectrum of particle  $i$ due to resonance decay is \cite{Kapusta:1977ce,Sollfrank:1990qz,Sollfrank:1991xm, Gorenstein:1987zm,Lo:2017sux}
\begin{equation}\label{eq:resonance_decay_spectrum}
\begin{split}
&E_i\frac{dN^{res}_i}{d^3\mathbf{p}_i} = \int_0^{\infty}dm \,W(m) \int \frac{d^3\mathbf{p}_{R}}{E_R}\left(E_R\frac{dN^{th}_R}{d^3\mathbf{p}_R}\right)\\
&\qquad\times\left[Br\times\left(\frac{1}{\Gamma_i}\right)\times\left(E_i\frac{d\Gamma_i}{d^3\mathbf{p}_i}\right)\right]\, .\\
\end{split}
\end{equation}
Here
the quantity in the square bracket
is the probability density for a resonance particle with energy and momentum $E_R$ and $\mathbf{p}_R$ decays to the particle $i$ which happens to have energy and momentum $E_i$ and $\mathbf{p}_i$. The  $Br$  is the branching ratio of the decay channel $Br=\Gamma_i/\Gamma_R$ with $\Gamma_R$ the total decay width of the resonance particle. The thermal spectrum of the resonance particle $E_RdN_R^{th}/d^3\mathbf{p}_R$ is computed by the Cooper-Frye formula Eq. \eqref{eq:cooper_frye}. The $W(m)$ integration describes the finite width effect of the resonance particle. The standard  functional form of $W(m)$ is chosen to be the Breit-Wigner distribution. In the zero width limit, it reduces to the Dirac delta function. For simplicity, we will chose $W(m) = \delta(m-m_R)$. Recent discussions on the finite width effects can be found in \cite{Huovinen:2016xxq,Vovchenko:2018fmh}. In the following, we will mainly focus on two-body decays. One crucial step is to compute the probability density associated with each decay channel. This may seem intimidating as each decay channel has its own decay matrix element $\mathcal{M}$ which in general depends on all the momenta of the particles involved in the decay process. Therefore, a rigorous treatment is to analyze the dynamics of each decay channel in detail to obtain the exact decay matrix element $\mathcal{M}$. A less rigorous but more practical approach, which has been adopted in several previous works, is to assume certain properties of the decay matrix element $\mathcal{M}$ to simplify the computation of the probability density. For example, the decay matrix $\mathcal{M}$ is assumed to be independent of momenta of the particles involved at the tree level. Under this assumption, the probability density can be solely represented by the phase space. Microscopic dynamics of the decay process completely drops out of the formula leaving only the phase space factor. With this assumption, analytic expressions of the probability density for two-body decays and three-body decays can be achieved. In this paper, we consider all the resonances included in the crossover EOS listed in \cite{Albright:2014gva}. Only two-body decays that directly produce protons will be considered in the estimation of the final proton spectrum.


\section{Longitudinal Dynamics of High Baryon Density Matter}
\label{sec:longitudinal_dynamics}

First of all, a few numerical setups and checks can be mentioned. The range of the spatial rapidity is chosen to be $-6\leq \eta \leq 6$. The space-time grids have spacings $\Delta \eta = 0.04$ and $\Delta \tau = 0.02\,\rm{fm/c}$. The free parameter $C_B$ in the baryon transport coefficient and the relaxation time is chosen to be $C_B =0.4$ for illustrative purposes.  In the numerical calculations, the energy conservation and the baryon conservation should be respected. The total energy defined by $E^{\rm{tot}} = \int T^{\mu 0}d^3\sigma_{\mu}$ with the hypersurface at constant proper time $\tau$ being $d^3\sigma_{\mu} = (\cosh{\eta}, 0, 0, -\sinh{\eta})\tau d\eta d^2\mathbf{x}_{\perp}$ has the explicit expression
\begin{equation}
E^{\rm{tot}}=S_A\int \tau d\eta\left(T^{\tau\tau}\cosh{\eta} + \tau T^{\tau\eta}\sinh{\eta}\right).
\end{equation}
Here $S_A$ is the transverse area. The total number of baryons defined by $N_B^{\rm{tot}} = \int J^{\mu}d^3\sigma_{\mu}$ has a contribution from the diffusive current
\begin{equation}\label{eq:check_total_baryons}
N_B^{\mathrm{tot}} = S_A\int \tau d\eta\,( n_Bu^{\tau} +  V^{\tau}). 
\end{equation}
In addition, the total entropy, which is defined by $S^{\rm{tot}} = \int s^{\mu}d^3\sigma_{\mu}$ with $s^{\mu} = su^{\mu} - \frac{\mu_B}{T}V^{\mu}$, should also be conserved for an ideal fluid and should not decrease for diffusive hydrodynamics. Its explicit expression is
\begin{equation}
S^{\mathrm{tot}} = S_A \int \tau d\eta \left[ su^{\tau} - \left(\frac{\mu_B}{T}\right)V^{\tau}\right].
\end{equation}

In the ideal fluid situation when the diffusive current is turned off $V^{\mu}=0$, the $E^{\mathrm{tot}}/S_A, N_B^{\mathrm{tot}}/S_A$ and $S^{\mathrm{tot}}/S_A$ are constants. Their initial values are $284.49\,\rm{GeV/fm}^2$, $4.2857\,\rm{fm}^{-2}$ and $339.16\,\rm{fm}^{-2}$, respectively. We check their values at each time step. Baryons are conserved to a very high precision. The total energy is conserved with less than $0.1\%$ numerical violations while the entropy is conserved with less than $0.2\%$ numerical violations. These tiny increases of the total energy and the total entropy are mainly due to the numerical viscosities of the Central-Upwind scheme in solving hydrodynamic equations. Further improvements of the algorithm can be found in \cite{KurganovLin:2007}. 

In the diffusive hydrodynamics with $V^{\mu}$ included, $E^{\mathrm{tot}}/S_A$ and $N_B^{\mathrm{tot}}/S_A$ are still expected to be constants while the entropy per transverse area $S^{\mathrm{tot}}/S_A$ is expected to increase.  The total energy in this situation is almost the same as in the perfect fluid situation. The total number of baryons, however, increases up to $2\%$ at the ending time of the hydrodynamic evolution. This $2\%$ percentage violation of baryon conservation is the accuracy of the numerical results on quantities related to baryons. There are several sources this small violation of baryon conservation can come from. After introducing the baryon diffusion current, the source terms Eqs.~\eqref{eq:physical_sources} contain time derivatives. This structure in principle violates the Central-Upwind scheme. The treatment of these time derivatives by the finite difference method in the codes could introduce uncertainties in the numerical results. In addition, the baryon diffusion current $V^{\mu}$ is theoretically expected to be smaller than the ideal part of the total baryon current $J^{\mu}_{\rm{id}}$. However, in numerically solving the baryon diffusion equation governing the time evolution of $V^{\mu}$, there is no small controlling parameter that guarantees $|V^{\mu}| \lesssim |J^{\mu}_{\rm{id}}|$. As a result, the $V^{\mu}$ can be comparable or larger than $J^{\mu}_{\rm{id}}$, especially in the spatial regions close to the edges of the system where the baryon densities are small. The baryon diffusion currents $V^{\mu}$ are regulated so as to make them smaller in these regions. 

\subsection{Baryon Diffusion}

\begin{figure}[t]
	\centering
	\includegraphics[scale=0.8]{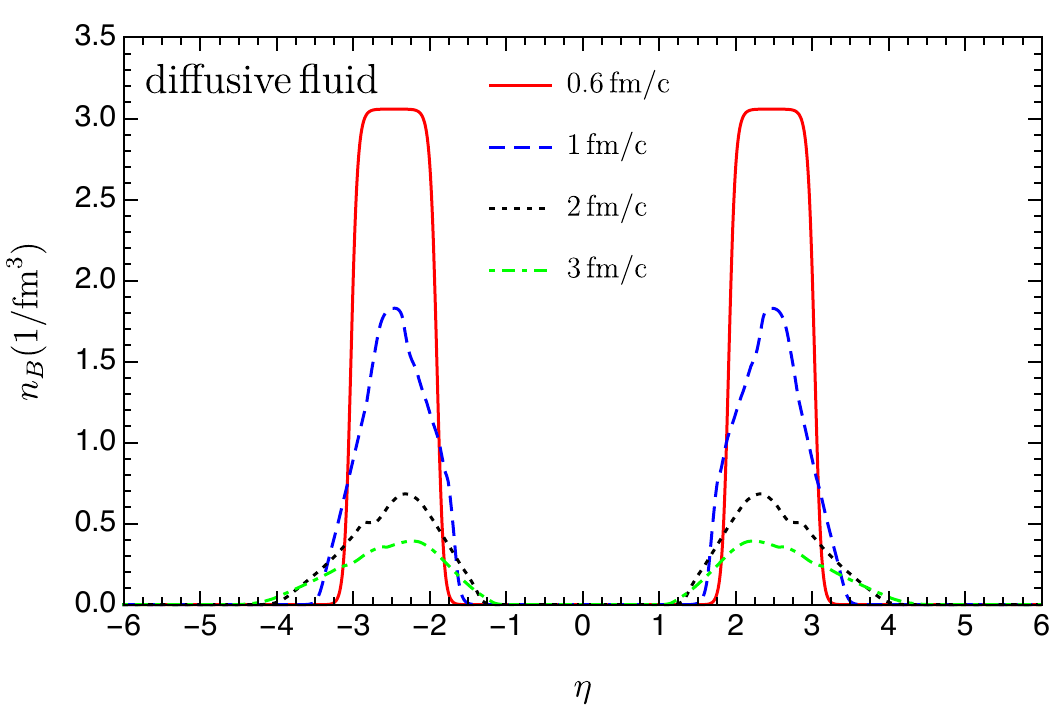}
	\caption{(color online) Profile of the proper baryon density $n_B(\eta)$ as it evolves with time hydrodynamically.}
	\label{fig:nB_eta_four_times_kinetic_nIF}
\end{figure}  

\begin{figure}[t]
	\centering
	\includegraphics[scale=0.8]{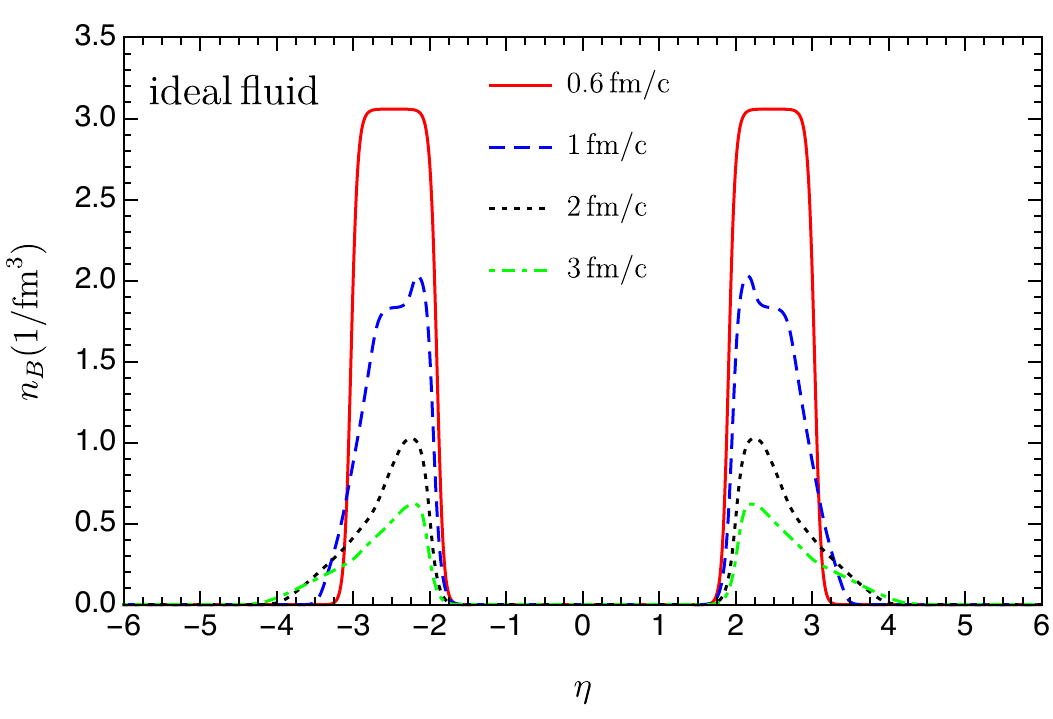}
	\caption{(color online) Profile of the proper baryon density $n_B(\eta)$ as it evolves with time according to the ideal fluid hydrodynamic equations.}
	\label{fig:nB_eta_four_times_ideal_nIF}
\end{figure} 
\begin{figure}[t]
	\centering
	\includegraphics[scale=0.8]{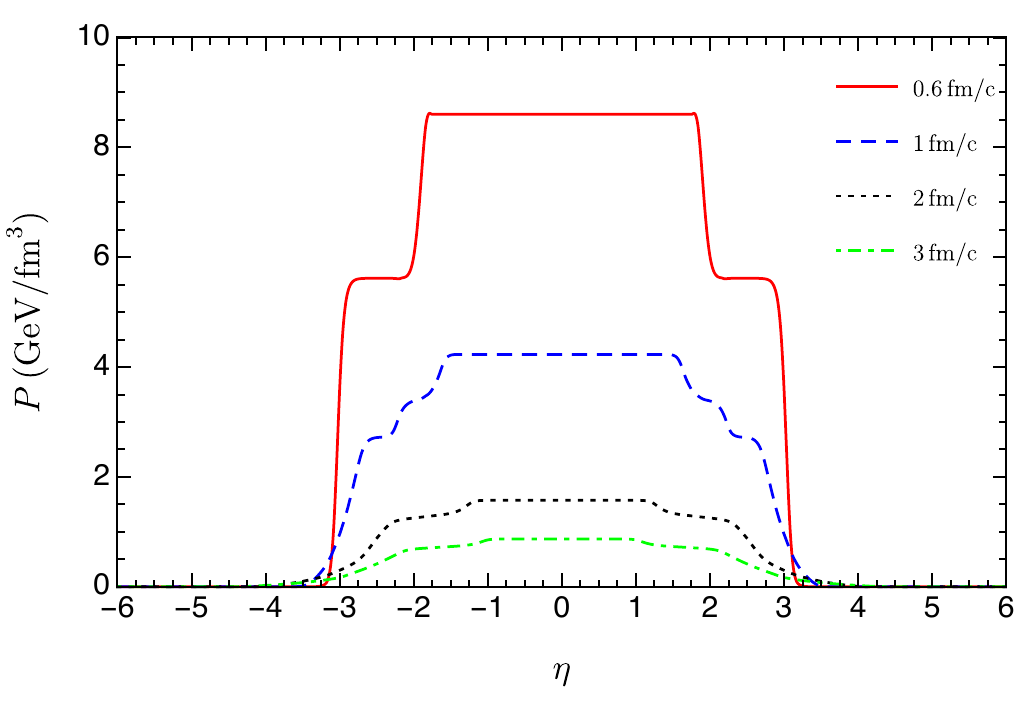}
	\caption{(color online) Profile of the pressure $P$ as it evolves with time according to the diffusive hydrodynamic equations.}
	\label{fig:pressure_eta_four_times_ideal_nIF}
\end{figure}

\begin{figure}[t]
	\centering
	\includegraphics[scale=0.8]{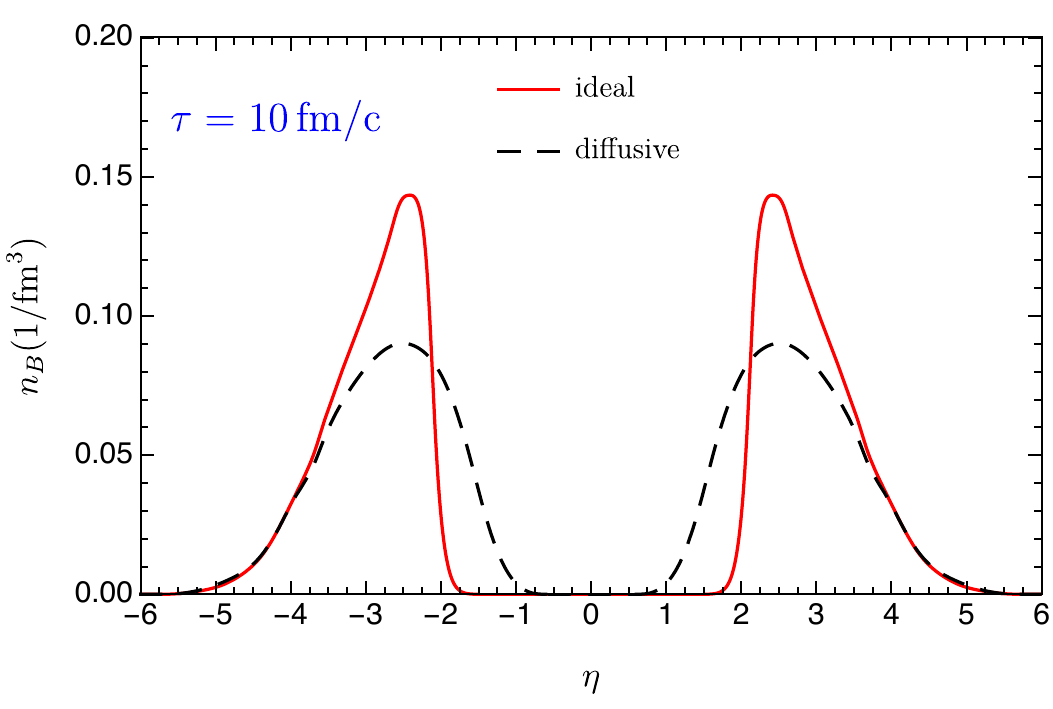}
	\caption{(color online) The proper baryon density distribution $n_B(\eta)$ at $\tau=10\,\rm{fm/c}$ from the ideal hydrodynamics and from the diffusive hydrodynamics.}
	\label{fig:nB_eta_ideal_diffusive_compare_nIF}
\end{figure}

 The proper baryon densities at $\tau=0.6$, $1.0$, $2.0$, $3.0$ $\,\rm{fm/c}$ are shown in Fig. \ref{fig:nB_eta_four_times_kinetic_nIF}.  Note that the area under the curve $n_B(\eta)$ is not the total number of baryons per unit area. The total number of baryons per unit area have to be computed using Eq. \eqref{eq:check_total_baryons}. As can be seen from Fig. \ref{fig:nB_eta_four_times_kinetic_nIF}, the magnitude of the proper baryon density decreases very fast at the early stage of the hydrodynamic expansion. It drops from the initial $3.0\,\rm{baryons/fm}^{3}$ at $\tau=0.6\,\rm{fm/c}$, which is around 20 times larger than the normal nuclear density, to around $0.4\,\rm{baryons/fm}^3$ at $\tau=3\,\rm{fm/c}$ which is less than three times larger than the normal nuclear density. As will be shown later, the total time of the hydrodynamic evolution is around $\tau_{\rm{tot}}\sim18\,\rm{fm/c}$. In Fig. \ref{fig:nB_eta_four_times_kinetic_nIF}, the baryon distribution spreads to both the smaller rapidity regions $|\eta|\lesssim 2$ and the larger rapidity regions $|\eta|\gtrsim 3$ as the system evolves. There are two factors that determine the change of the baryon distribution during the hydrodynamic evolution. One is the longitudinal hydrodynamic expansion driven by the gradient of the pressure. The other is the baryon diffusion driven by the gradient of the fugacity. The effects of the longitudinal hydrodynamic expansion push the baryons to larger spatial rapidity regions as shown in Fig. \ref{fig:nB_eta_four_times_ideal_nIF} where the baryon diffusion current is turned off: $V^{\mu}=0$. No baryons spread to the central region in the ideal fluid dynamical evolution. The pressure profiles are shown in Fig. \ref{fig:pressure_eta_four_times_ideal_nIF}. In the ideal fluid, the change of baryon density is governed by $D n_B = - n\theta$ and $ Du^{\mu} =\nabla^{\mu}P/(\varepsilon+P)$. A monotonically decreasing pressure profile from the central region to the forward/backward rapidity regions causes the expansion. Figure \ref{fig:nB_eta_ideal_diffusive_compare_nIF} shows the proper baryon density distributions at $\tau=10\,\rm{fm/c}$ from the ideal hydrodynamics and the diffusive hydrodynamics. Diffusion causes the baryons to be transported from the backward/forward rapidity regions to the central region. However, baryon diffusion to larger rapidity regions $|\eta|\gtrsim 3$ are suppressed. The spreading of baryons to the $|\eta|\gtrsim 3$ regions are mainly caused by the hydrodynamical expansion. Close to the edge of the system $|\eta|\gtrsim 3.8$, the two distributions overlap because of the regulation of the baryon diffusion current $V^{\mu}$ in these regions. 
\begin{figure}[t]
	\centering
	\includegraphics[scale=0.85]{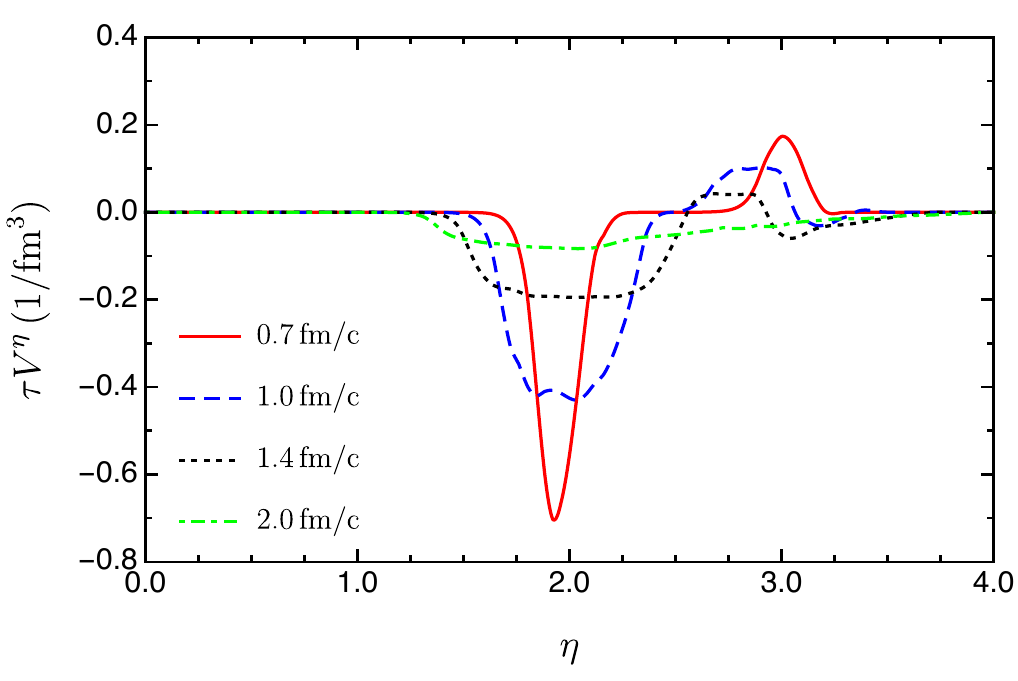}
	\caption{(color online) The baryon diffusion current $\tau V^{\eta}$ as it evolves with time hydrodynamically. Only the postive rapidity region is shown.}
	\label{fig:tauVeta_kinetic_nIF}
\end{figure} 

\begin{figure}[t]
	\centering
	\includegraphics[scale=0.75]{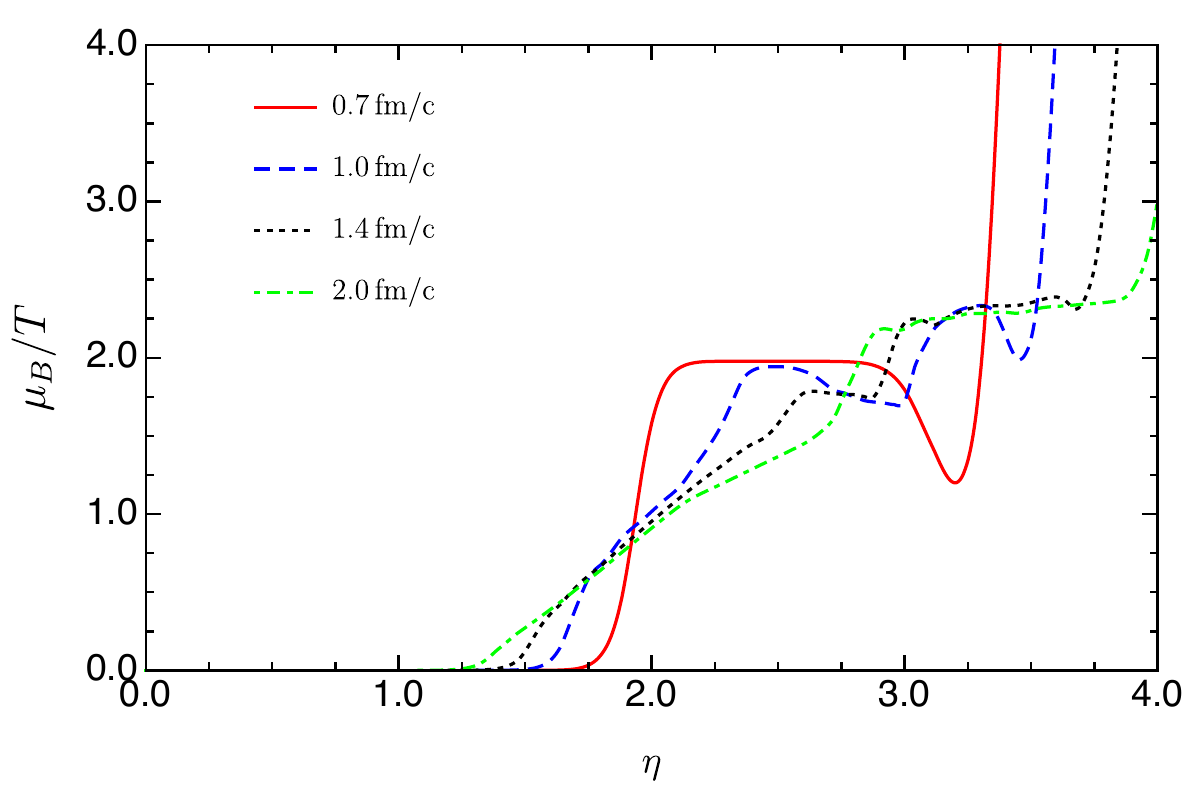}
	\caption{(color online) The $\mu_B/T$ profile evolves with time hydrodynamically. Only the positive rapidity region is shown.}
	\label{fig:muB_over_T_kinetic_nIF}
\end{figure}

The dynamics of the baryon diffusion can also be further understood by looking at the baryon diffusive current in detail. The baryon diffusion currents at four different proper times $\tau = 0.7, 1.0, 1.4, 2.0\,\rm{fm/c}$ are plotted in Fig. \ref{fig:tauVeta_kinetic_nIF}. The baryon diffusion current is assumed to be zero at the initial time $\tau=0.6\,\rm{fm/c}$.   The direction and magnitude of the baryon diffusion are determined by the fugacity gradient. In the relativistic Navier-Stokes limit, we have $V^{\mu} = \kappa_B\nabla^{\mu}(\frac{\mu_B}{T})$. Therefore, in the 1+1D situation
\begin{equation}
\begin{split}
\tau V^{\eta} &=V^{z}\cosh{\eta} - V^{t}\sinh{\eta}\\
&\sim -\kappa_B\cosh{(y-\eta)}\left[ \tau u^{\eta}\partial_{\tau} + u^{\tau}\tau^{-1}\partial_{\eta}\right]\left(\frac{\mu_B}{T}\right)\\
&\sim -\kappa_B\left[\cosh{(y-\eta)}\right]^2\frac{1}{\tau}\partial_{\eta}\left(\frac{\mu_B}{T}\right)
\end{split}
\end{equation}
where in the last line we only focus on the spatial gradient term. The baryon diffusion current $\tau V^{\eta}$ is inversely proportional to the rapidity gradient of $\mu_B/T$ and is scaled by $1/\tau$. The factor $1/\tau$, which is the same as the expansion rate $\partial_{\mu}u^{\mu}$ in the boost-invariant case, indicates that the magnitude of $\tau V^{\eta}$ is larger at early times and becomes smaller and smaller at late times due to the hydrodynamical expansion. The explicit profiles of $\mu_B/T$ as a function of $\eta$ are shown in Fig. \ref{fig:muB_over_T_kinetic_nIF}. Around $\eta\sim 2$, the value of $\mu_B/T$ increases from zero to about two at $\tau=0.7\,\rm{fm/c}$. This is a general feature of the space-time picture of high energy heavy-ion collisions where very few baryons are left in the central region and most of the baryons are carried away by the nuclear remnants. Taking a derivative of $\mu_B/T$, one gets the ``plateau-valley-plateau" structure in the $\tau V^{\eta}$ profile in the connecting area between the central region and the fragmentation regions as shown in Fig. \ref{fig:tauVeta_kinetic_nIF}. We want to emphasize again that this is a general feature of high energy heavy-ion collisions. In the extreme case of low collision energy where the two colliding nuclei are completely stopped, baryons are concentrated in the central region. The $\mu_B/T$ instead should be decreasing from the central region to the fragmentation regions. As a consequence, one gets the ``plateau-hill-plateau" structure in the baryon diffusion current $\tau V^{\eta}$ which indicates that the baryons diffuse from the central region to the fragmentation regions. For the $\mu_B/T$ profiles shown in Fig. \ref{fig:muB_over_T_kinetic_nIF}, as baryons diffuse from the fragmentation regions to the central regions, the $\mu_B/T$ distribution is lowered and spread around $\eta \sim 2$.  The deep valley in the $\tau V^{\eta}$ profile correspondingly becomes shallower and more flat. The region around $\eta\sim 3$ in Fig. \ref{fig:muB_over_T_kinetic_nIF} needs more explanation. This is the connecting area between the high baryon density matter and the vacuum, which is different from the connecting area around $\eta \sim 2$ where high baryon density matter and  the QGP with zero baryon densities are connected. The value of $\mu_B/T$ first decreases and then quickly increases to infinity around $\eta \sim 3$ at $\tau=0.7\,\rm{fm/c}$. The values of $\mu_B$ and $T$ are determined by the equation of state, given the values of the energy density $\varepsilon$ and the baryon density $n_B$. Around $\eta\sim 3$, both $\varepsilon$ and $n_B$ are small while around $\eta\sim 2$, the $n_B$ is small but $\varepsilon$ is very large. The feature of $\mu_B/T$ around $\eta\sim 3$ thus reflects the property of the equation of state at small $n_B$ and $\varepsilon$. See also similar structure using a different equation of state in Ref. \cite{Denicol:2018wdp}. This particular feature of $\mu_B/T$ results in a small hill structure around $\eta\sim  3$ in the $\tau V^{\eta}$ profile as shown in Fig. \ref{fig:tauVeta_kinetic_nIF}. Baryons thus diffuse from the fragmentation regions to even larger rapidity regions where the vacuum locates. Gradually, this small hill structure in $\tau V^{\eta}$ around $\eta \sim 3$ is flattened as the system expands and the baryons diffuse; we are left with a monotonic increasing rapidity profile of $\mu_B/T$. 

\begin{figure}[t]
	\centering
	\includegraphics[scale=0.75]{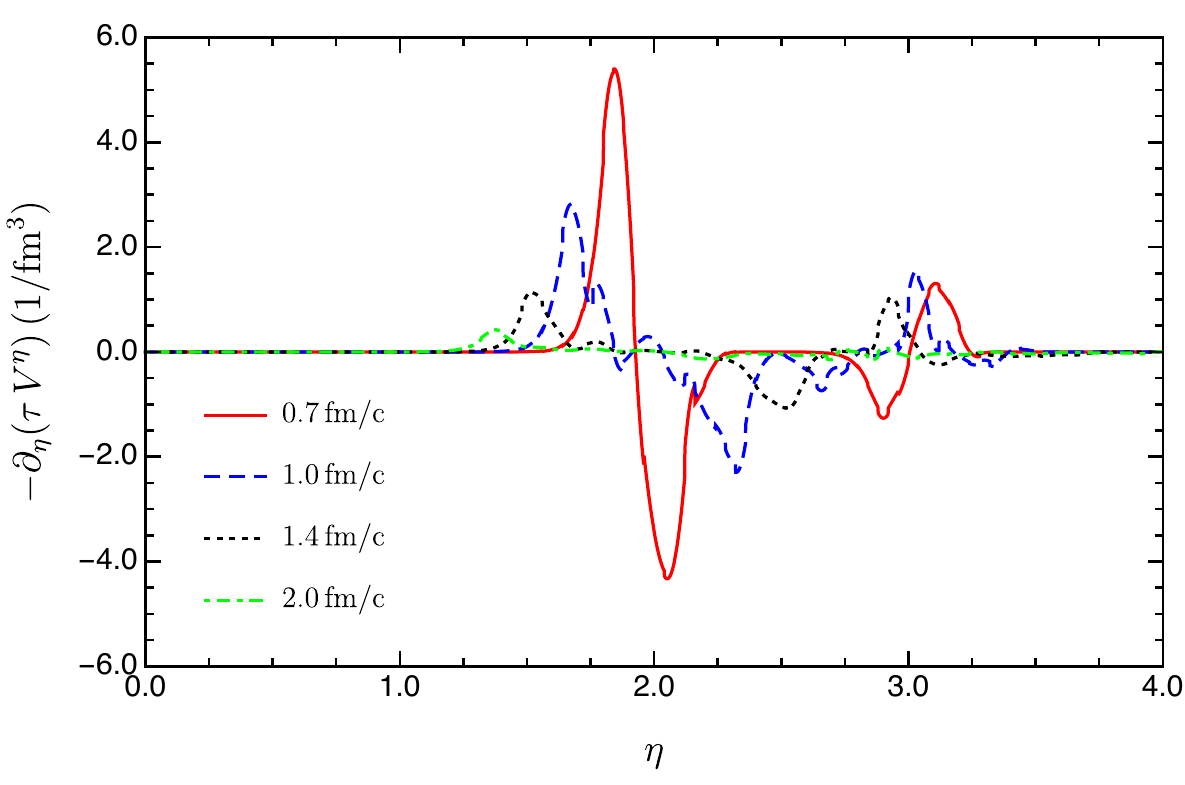}
	\caption{(color online) The gradient of baryon diffusion current $-\partial_{\eta}(\tau V^{\eta})$ as it evolves with time hydrodynamically. Only the postive rapidity region is shown.}
	\label{fig:derivative_tauVeta_kinetic_nIF}
\end{figure}

The reason why baryons diffuse from the fragmentation regions to the central region given the ``plateau-valley-plateau" structure of the baryon diffusion current $\tau V^{\eta}$ are explicitly shown in Fig. \ref{fig:derivative_tauVeta_kinetic_nIF}. The source term $S_B$ in the baryon conservation equation Eq. \eqref{eq:1d_explicit} has a rapidity gradient term $S_B \sim -\partial_{\eta}(\tau V^{\eta})$ as shown in Eq. \eqref{eq:physical_sources}. The change of the proper baryon density $n_B$ after one time step due to the rapidity gradient of the baryon diffusion current can be approximated as
\begin{equation}
\Delta n_B \sim -\frac{1}{u^{\tau}} \frac{\Delta \tau}{\tau} \partial_{\eta} (\tau V^{\eta}). 
\end{equation}
Therefore, the ``plateau-valley-plateau" structure of the $\tau V^{\eta}$ causes the baryons to increase on one side of the valley and the baryons to decrease on the other side of the valley. Overall, one sees the baryons are transported from the fragmentation regions to the central region. Likewise, as the system expands and the baryons diffuse, the deep valley becomes more flat. Baryon diffusion still happens on the edge of the valley but the magnitudes are reduced due to the expansion of the system and the shallowness of the valley, as shown in Fig. \ref{fig:derivative_tauVeta_kinetic_nIF}.

\subsection{Rapidity Dependent Observables}
In the numerical calculations, we chose the freezeout energy density to be $\varepsilon_{FO} = 0.5\,\rm{GeV/fm}^3$ and found that the whole system freezes out at $\tau = 17.4\,\rm{fm/c}$ . Particle momentum space distributions can then be computed by the Cooper-Frye formula and by including contributions from the resonance decays. We focus on particles that carry baryon charge. Neutral particles like neutrons are difficult to measure experimentally. The BRAHMS collaboration has measured the proton, antiproton and net-proton rapidity distribution $dN/dy$ for $y$ up to $3.1$ for Au+Au collision at $\sqrt{s_{NN}} = 200\,\rm{GeV}$ in the $0-5\%$ centrality bin \cite{Bearden:2003hx}. Our calculation is not the full 3+1D, thus detailed quantitative comparisons with the experimental data  are not practical. However, the experimentally measured rapidity distributions of proton, antiproton and net-proton can put a qualitative constrain on the longitudinal dynamics of the high baryon density matter predicted by the 1+1D diffusive hydrodynamics. 
\begin{figure}[t]
	\centering
	\includegraphics[scale=0.8]{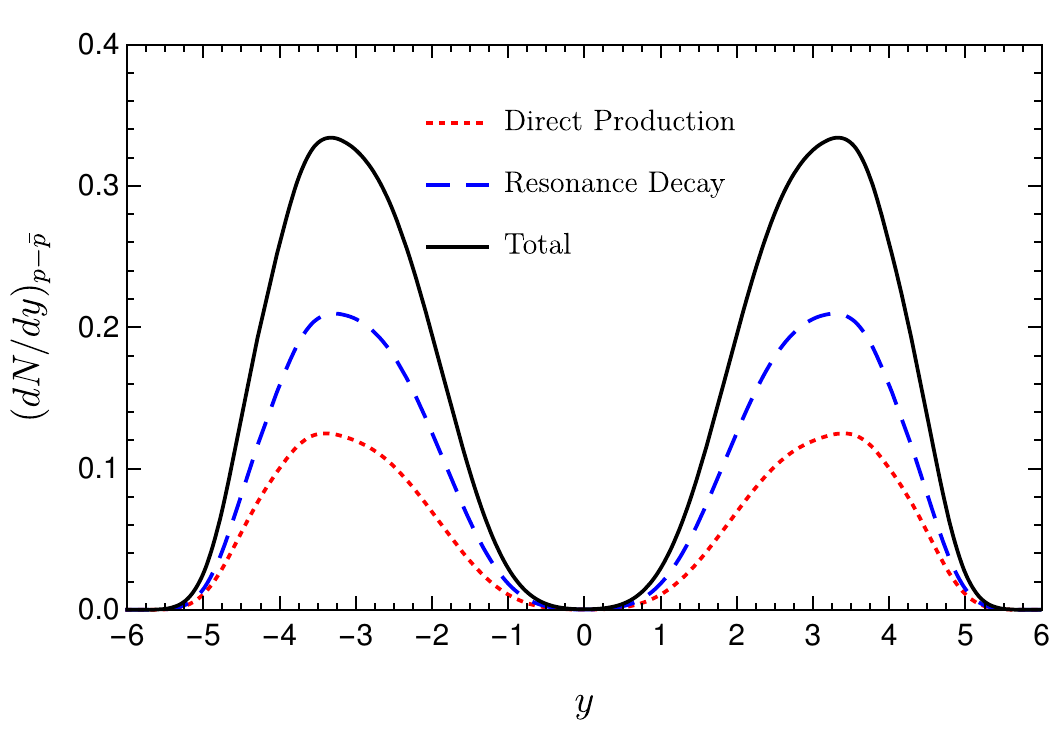}
	\caption{(color online) Comparison of the net-proton rapidity distributions from direct thermal production and resonance decays.}
	\label{fig:dNdy_netp_thermal_resonance}
\end{figure} 

\begin{figure}[t]
	\centering
	\includegraphics[scale=0.8]{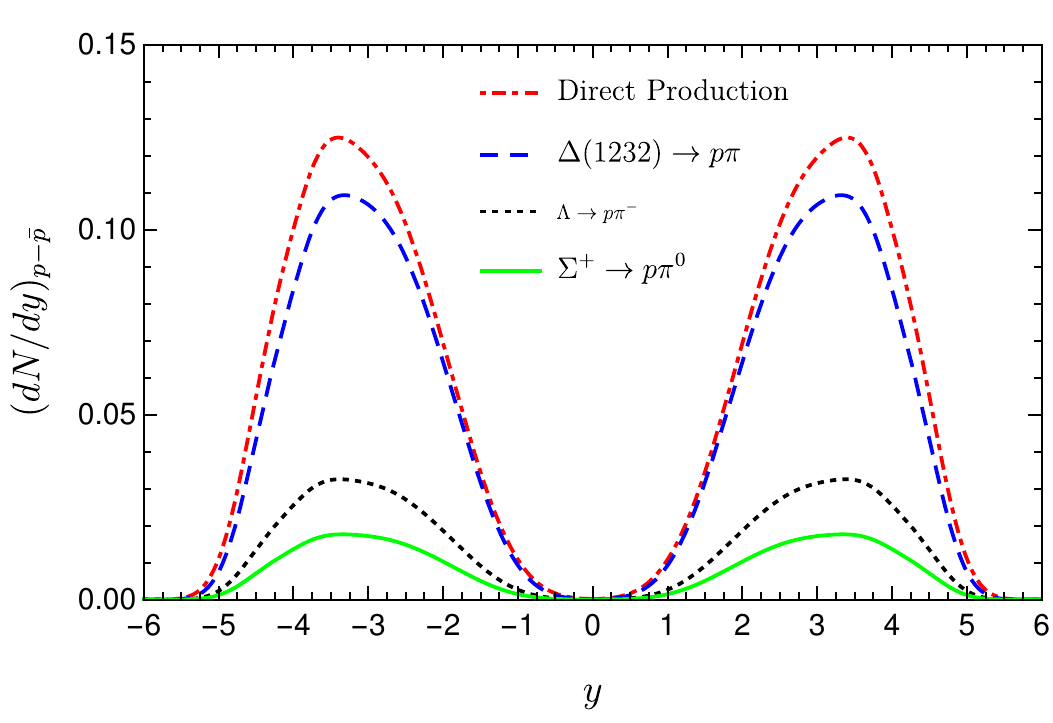}
	\caption{(color online) Comparison of the net-proton rapidity distributions from direct thermal production and three resonance decays $\Delta(1232)\rightarrow p\pi$, $\Lambda\rightarrow p\pi^-$ and $\Sigma^+\rightarrow p \pi^0$.}
	\label{fig:dNdy_netp_channels}
\end{figure} 

Figure \ref{fig:dNdy_netp_thermal_resonance} shows the net-proton rapidity distribution and the contributions from direct thermal production and from resonance decays. Resonance decays contribute more on the production of proton than the direct thermal emission. Including the resonance decays increases the magnitude of the net-proton rapidity distribution while the shape of the rapidity distribution remains unchanged. It would be crucial to include the resonance decays when one compares theoretical predictions with experimental data. Note that the $dN/dy$ has unit of  $1/\rm{fm}^2$, as we consider a unit transverse area around the central core of the nuclei. Bulk variables are counted as per unit area. Figure \ref{fig:dNdy_netp_channels} shows the net-proton rapidity distributions from three different resonance decays. The resonances $\Delta(1232)$, $\Lambda$ and $\Sigma^+$, whose masses are $m_{\Delta} = 1232\,\rm{MeV}$, $m_{\Lambda} = 1115.7\,\rm{MeV}$ and $m_{\Sigma} = 1189.4\,\rm{MeV}$, are the baryons that have the lowest masses besides the proton and neutron. The $\Delta(1232)\rightarrow p\pi$ actually consists of three channels $\Delta^{++}(1232)\rightarrow p\pi^+$, $\Delta^+(1232)\rightarrow p\pi^0$ and $\Delta^0(1232)\rightarrow p\pi^-$. 

\begin{figure}[t]
	\centering
	\includegraphics[scale=0.8]{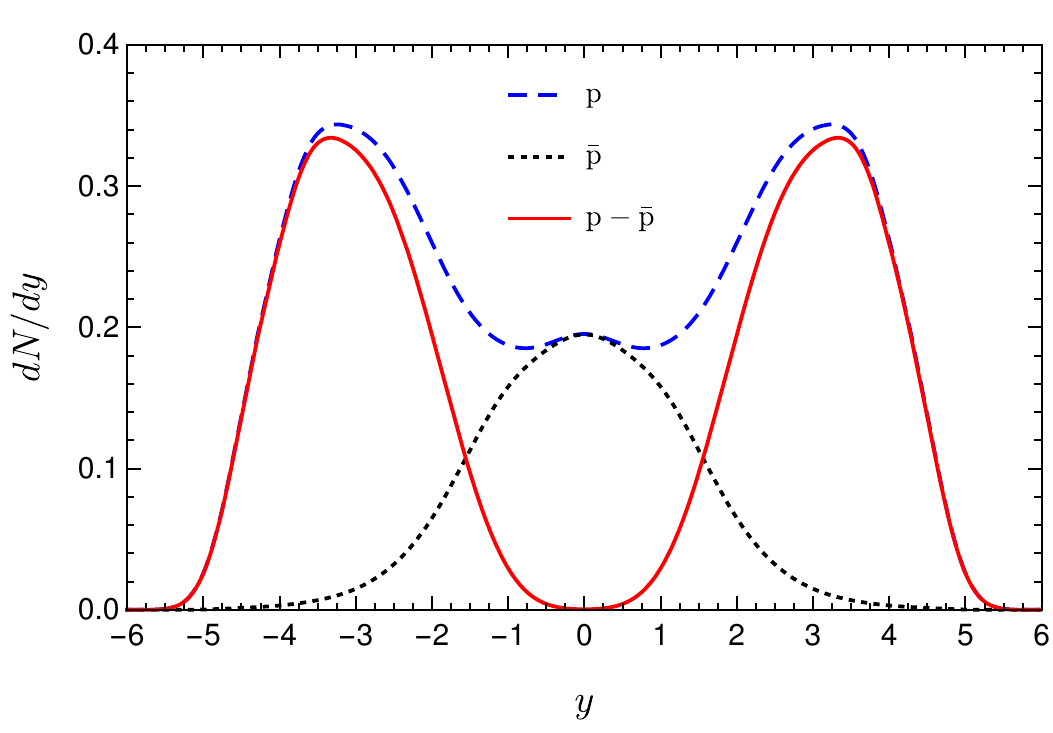}
	\caption{(color online) The proton, antiproton and net-proton rapidity distribution after the freezeout. Both the direct thermal production and contributions from the resonance decays are included.}
	\label{fig:dNdy_p_pbar_netp}
\end{figure} 
\begin{figure}[t]
	\centering
	\includegraphics[scale=0.8]{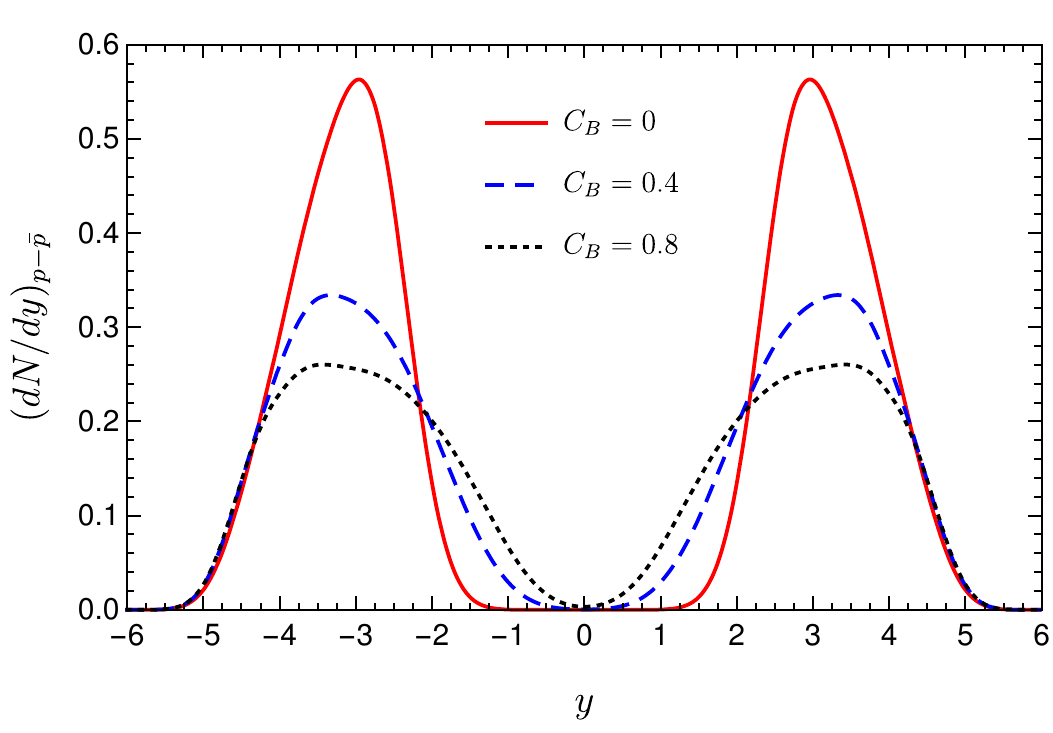}
	\caption{(color online) The net-proton distributions after freezeout for three different values of $C_B$.}
	\label{fig:dNdy_three_CBs}
\end{figure}

As shown in Fig. \ref{fig:dNdy_p_pbar_netp}, the net-proton distribution has a double peak structure which is in qualitative agreement with experimental findings. The difference lies in the valley between the two peaks around $y\sim 0$. Experimental data reveals that there is a small amount of baryon charge residing in the central region with the baryon chemical potential about $25\,\rm{MeV}$ at chemical freezeout \cite{Andronic:2012dm}. In the experimental data, the net-proton rapidity distribution has nonzero values around $y\sim 0$ and the valley between the two peaks is found to be shallower and more flat. As a consequence, this small baryon chemical potential around $y\sim 0$ predicts slightly more protons than the antiprotons. In Fig. \ref{fig:dNdy_p_pbar_netp}, the $(dN/dy)_p$ and $(dN/dy)_{\bar{p}}$ are the same around $y\sim 0$ because $\mu_B\simeq 0$ there. Antiprotons do not exist before the collision of the two nuclei, they come from pair production associated with the protons or from hadronization. Finally, the baryons are distributed as Dirac delta functions around $y= \pm 2.47$ in the momentum space at the initial time $\tau=0.6\,\rm{fm/c}$ . Hydrodynamic expansions and baryon diffusions spread the net baryon distribution over an wide range of $y$ in the momentum space as illustrated in Fig. \ref{fig:dNdy_p_pbar_netp}.

Where does the small amout of net baryons at $y\sim 0$ come from? Are they already there at the initial time $\tau = 0.6\,\rm{fm/c}$ because the collision is not fully transparent even at $\sqrt{s_{NN}}=200\,\rm{GeV}$? Or do the baryons come from diffusion from the fragmentation regions to the central region during the hydrodynamic evolution? Our modeling and calculations cannot evaluate the first possibility as full transparency was assumed. The second possibility would require an extraordinary strength of baryon diffusion. Given the initial conditions, the strength of the baryon diffusion is controlled by the baryon transport coefficient $\kappa_B$. However,  $\kappa_B$ can not be arbitrarily large as the requirement $|V^{\mu}| \lesssim |J^{\mu}_{\rm{id}}|$ has to be satisfied, otherwise the diffusive hydrodynamics breaks down.  Figure \ref{fig:dNdy_three_CBs} shows the net-proton distributions after the freezeout for three different values of $\kappa_B$ (the $C_B$ is a prefactor in the expression of $\kappa_B$ in Eq.\eqref{eq:kappaB_kinetic_theory}). As one increases the value of $\kappa_B$, more baryons diffuse to the central region. However, we have checked that a very large value $C_B =2$ still can not reproduce the expected net baryons at $y\sim 0$. Also the hydrodynamic code becomes unstable. Our calculations therefore do not favor the second possibility.

\begin{figure}[t]
	\centering
	\includegraphics[scale=0.65]{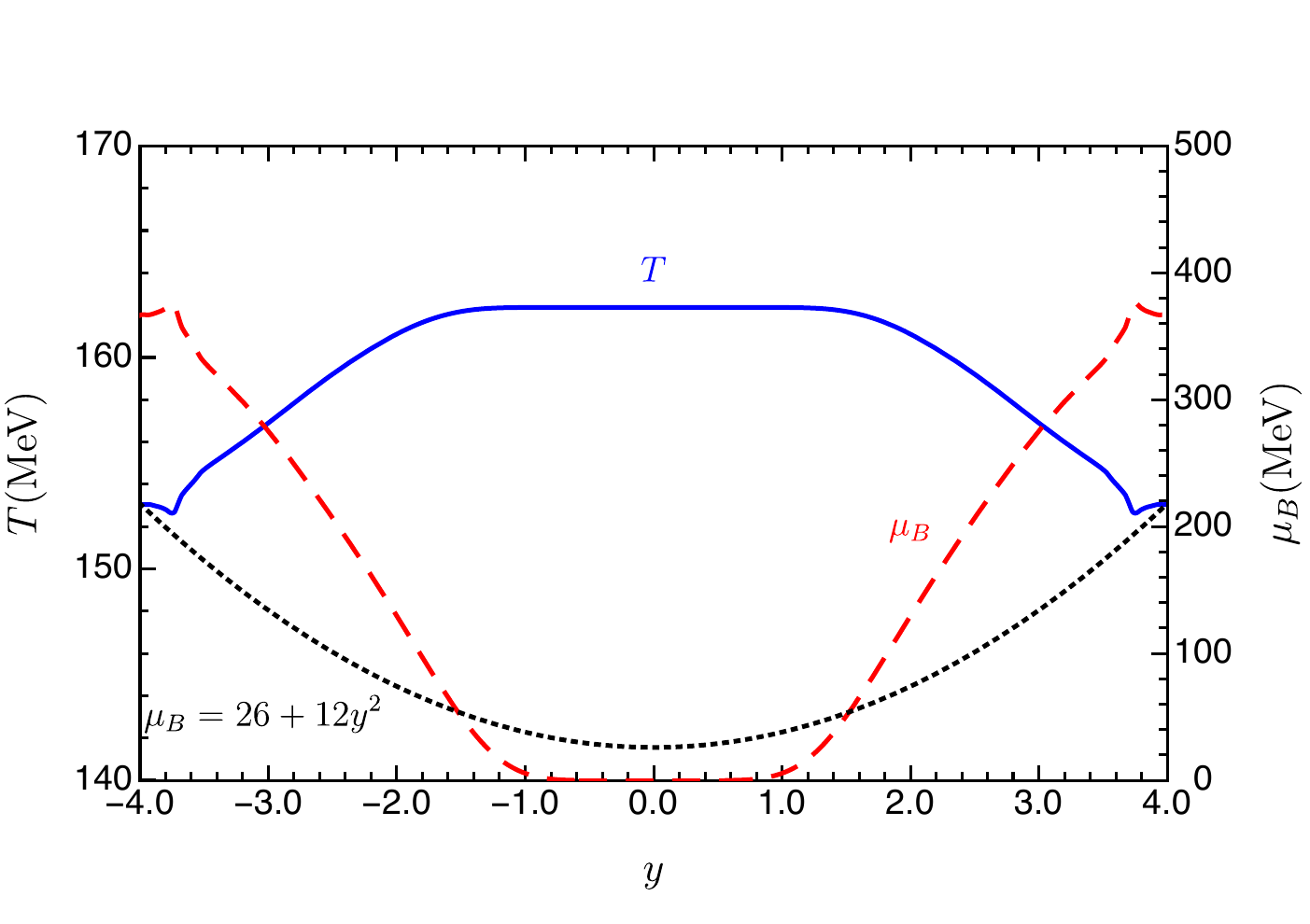}
	\caption{(color online) The temperature $T$ and the baryon chemical potential $\mu_B$ as functions of rapidity $y$ at the Cooper-Fye freezeout. The form $\mu_B = 26 + 12 y^2$ from Ref. \cite{Becattini:2007qr} is plotted for reference. }
	\label{fig:temp_muB_vs_y_freezeout}
\end{figure}

Figure \ref{fig:temp_muB_vs_y_freezeout} shows the temperature and the baryon chemical potential as functions of the rapidity $y$ at the Cooper-Frye freezeout. The values of temperature and baryon chemical potential at the chemical freezeout can be reconstructed from the thermal statistical model in fitting hadron yields and hadron yield ratios. An attempted extraction of the rapidity dependence of $T$ and $\mu_B$ for Au+Au collision at $\sqrt{s_{NN}} = 200\,\rm{GeV}$ using the data from the BRAHMS collaboration is presented in Ref. \cite{Becattini:2007qr}, where the parabolic form of the baryon chemical potential $\mu_B = 26 + 12 y^2$ was assumed. Again, our results show the baryon chemical potentials are zero around $y\sim 0 $ while the parabolic parameterization has $\mu_B = 26\,\rm{MeV}$ there. At the moment of freezeout, from the central region to the fragmentation regions, the baryon chemical potential increases while the temperature decreases. The distribution of $T$ and $\mu_B$ as functions of rapidity $y$ indicate that a rapidity scan would cover a wide range of $(T,\mu_B)$ points on the QCD phase diagram and can be used to search for the critical point in high energy heavy-ion collisons, which is different from the low energy Beam Energy Scan.

\begin{figure}[t]
	\centering
	\includegraphics[scale=0.8]{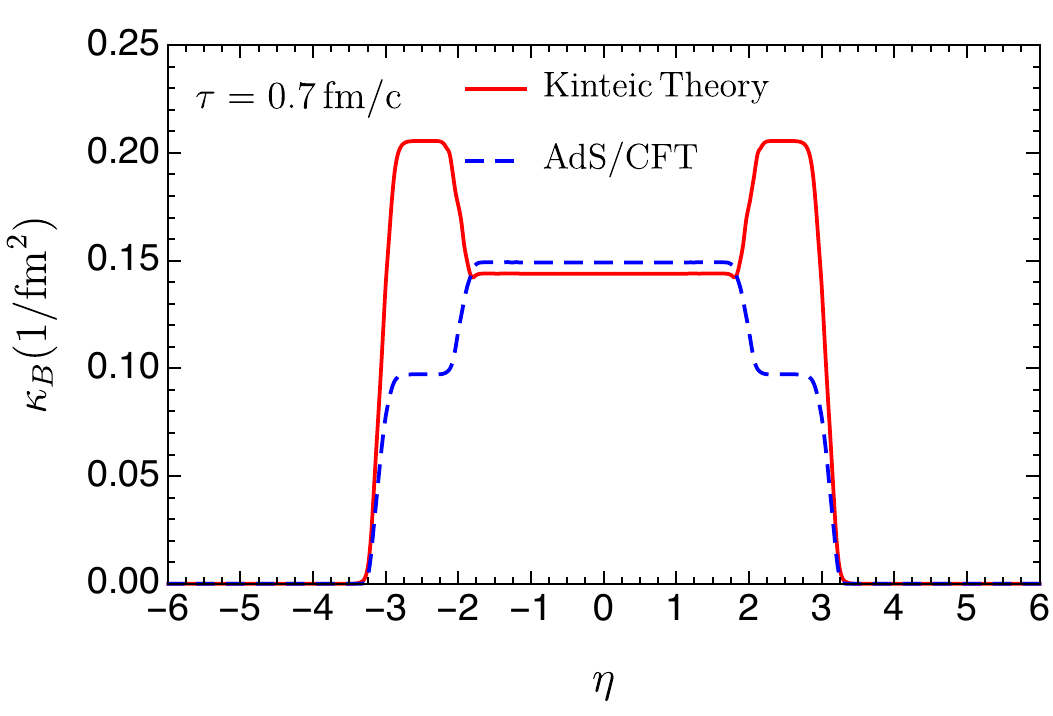}
	\caption{(color online) The initial profiles of the baryon transport coefficient $\kappa_B$ obtained from the kinetic theory approach and the AdS/CFT approach. }
	\label{fig:kB_vs_eta_compares}
\end{figure} 

\begin{figure}[t]
	\centering
	\includegraphics[scale=0.8]{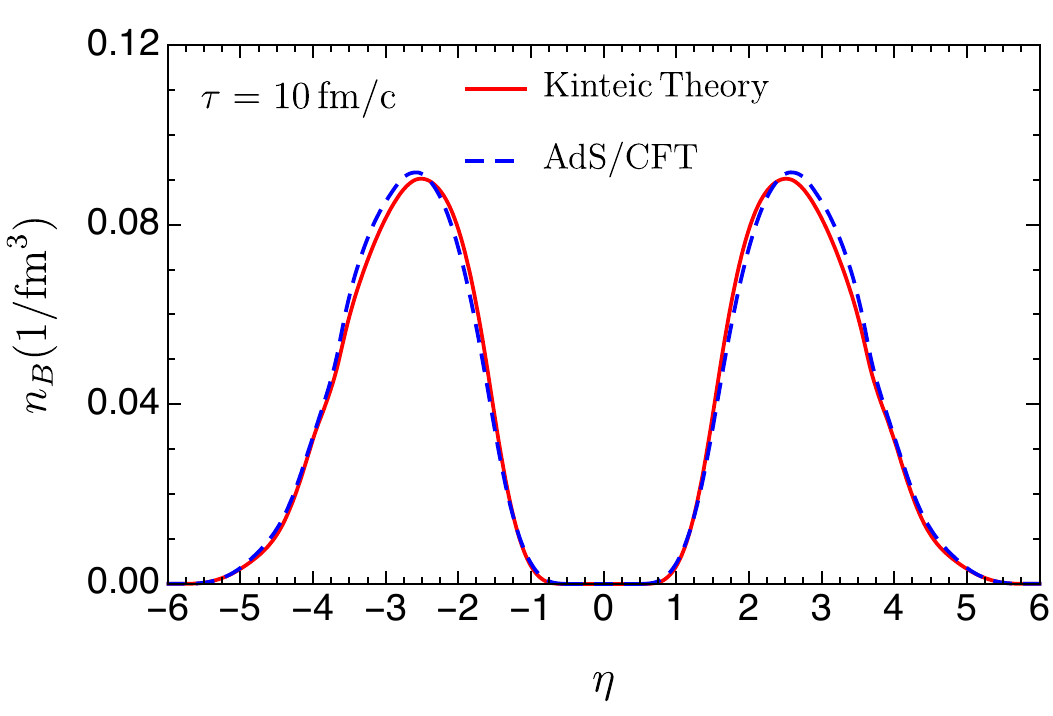}
	\caption{(color online) The baryon density spatial distribution at late time of the hydrodynamic evolution from two different baryon transport coefficients $\kappa_B$. }
	\label{fig:nB_vs_eta_compare_kappaB}
\end{figure} 

\begin{figure}[t]
	\centering
	\includegraphics[scale=0.8]{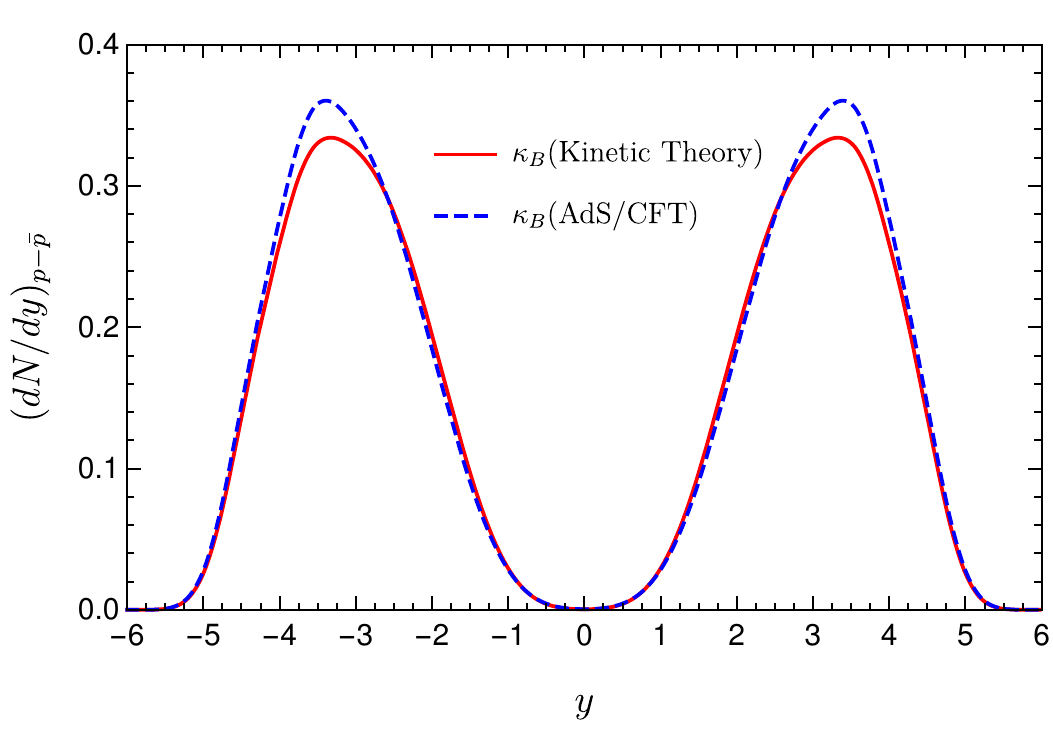}
	\caption{(color online) The net-proton rapidity distributions after freezeout from two different baryon transport coefficients $\kappa_B$. }
	\label{fig:dNdy_vs_y_compare_kappaB}
\end{figure} 

\subsection{Comparing Different $\kappa_B$}
The baryon transport coefficient $\kappa_B$ is not well known. In the above numerical results, we used the expression of $\kappa_B$ from Eq. \eqref{eq:kappaB_kinetic_theory} which was derived using the kinetic theory approach. In this section, we will compare numerical results from using a different baryon transport coefficient Eq. \eqref{eq:kappaB_holographic}, which was suggested by the AdS/CFT approach.  The $\kappa_B$ from the kinetic theory approach is not completely determined by the thermodynamic quantities as its magnitude is controlled by a free parameter $C_B$. The $\kappa_B$ from the AdS/CFT approach, however, is completely fixed by the thermodynamic state.  In the comparison, we assume the relaxation time is the same $\tau_V = C_B/T$ and we pick $C_B=0.4$. Figure \ref{fig:kB_vs_eta_compares} shows the two baryon transport coefficients as functions of $\eta$ at the initial time $\tau = 0.7\,\rm{fm/c}$. Both $\kappa_B$ have similar values in the limit of vanishing baryon density.  The $\kappa_B$ from the kinetic theory approach is larger than that from the AdS/CFT approach at non-zero baryon densities. As a consequence, the baryon diffusion is stronger when using the $\kappa_B$ from the kinetic theory approach. However, Fig \ref{fig:nB_vs_eta_compare_kappaB} shows the baryon density distributions at late times of the hydrodynamic evolution are almost the same for the two $\kappa_B$. This is easy to understand as the baryon diffusion is more efficient at early times of the hydrodynamic evolution. At late times, hydrodynamic expansion plays a  dominant role, leaving the differences resulting from the two different $\kappa_B$ very small. Figure \ref{fig:dNdy_vs_y_compare_kappaB} presents the net-proton rapidity distribution after freezeout using the two different $\kappa_B$. The $(dN/dy)_{p-\bar{p}}$ has a slightly larger peak value when using $\kappa_B$ from the AdS/CFT approach. Other than that, the two distributions overlap with each other. Although these two baryon transport coefficients come from completely different  approaches, the physics at late times of the hydrodynamic evolution is not very sensitive to their differences.

\section{Conclusion and Outlook}
In this paper, we have performed numerical calculations of the 1+1D diffusive hydrodynamic equations given the high baryon density initial conditions for Au+Au central collision at $\sqrt{s_{NN}} = 200\,\rm{GeV}$. We gave detailed discussions on the dynamics of baryon diffusion. Baryons are found to diffuse from the fragmentation regions to the central region due to the fugacity gradient which overcomes the counter-force of hydrodynamic expansion. The number of baryons transported to the central region is controlled by the baryon transport coefficient. It is difficult to achieve the observed number of baryons in the central region by purely baryon diffusion alone because the baryon transport coefficient can not be arbitrarily large before hydrodynamics break down. The proton, antiproton and net-proton rapidity distributions after the freezeout are in qualitative agreement with experimental findings. At freezeout, the baryon chemical potential increases from $0$ up to $400\,\rm{MeV}$ while the temperature decreases from about $165\,\rm{MeV}$ to $150\,\rm{MeV}$ as the rapidity increases from $y=0$ to $y=4$. These values reinforce the idea of a rapidity scan in high energy heavy-ion collisions to explore the QCD phase diagram. 

Unfortunately, experimental measurements of particle production at large rapidity are very challenging in the collider mode. The BRAHMS collaboration at RHIC had the capability of measuring (anti)protons, kaons and pions up to rapidity $y\sim 3.1$. A detailed rapidity scan has not been performed yet.  It is desired that experimentalists can identify particles' identities (thus knowing their masses) and measure their transverse momenta while scanning through momentum space rapidity. A more promising approach in measuring the fragmentation regions could be fixed target experiments. For example, the AFTER@LHC project \cite{Rakotozafindrabe:2012ei, Brodsky:2012vg}, a fixed target experiment using the $2.76\,\rm{TeV}$ Pb beam which is equivalent to the $\sqrt{s_{NN}} = 72\,\rm{GeV}$  Pb+Pb collision in the center-of-momentum frame, is under study. Two recent studies on the idea of a rapidity scan in the AFTER@LHC project can be found in Refs. \cite{Karpenko:2018xam,Begun:2018efg}. In a fixed target experiment, detectors are located in the forward rapidity region behind the target in the Lab frame. This forward rapidity region in a fixed target experiment corresponds to one of the fragmentation regions in the collider mode. It is worth emphasizing that the major difficulty in measuring the large rapidity regions is particle identification (PID). In design the BRAHMS detector, the largest rapidity $y=4$ is equivalent to $2^{\circ}$ away from the beam direction. With particles produced in directions so close to the beam direction, signals are easily buried in the original beam particles. It is also worth noting that there are dedicated experiments, the LHCf \cite{Adriani:2015nwa} and the RHICf \cite{Itow:2014oea} experiments, that measure the neutrons in the forward direction.  Last but not least, it is interesting that  a rapidity scan around the central rapidity might be indispensible in pinning down the critical point even in the low energy BES program \cite{Brewer:2018abr}.  

In the future, numerical calculations of the full 3+1D hydrodynamics including both viscous and diffusive effects, together with the high baryon density initial conditions, are needed to understand the three dimensional dynamics of baryon diffusions and to quantitatively compare with experimental data. 

\section*{Acknowledgement}

We are grateful to M. Albright for sharing his codes of the crossover EoS. We thank J. Kapusta for carefully reading the manuscript and his enlightening comments. We thank C. Plumberg for helpful discussions. We would also like to acknowledge computational support from the University of Minnesota Supercomputing Institute.  M. Li was supported by the Doctoral Dissertation Fellowship of the University of Minnesota and the U. S. Department of Energy grant  DE-FG02-87ER40328. C. Shen was supported by the U.S. Department of Energy, Office of Science, Office of Nuclear Physics, under Contract No. DE-SC0012704 and within the framework of the Beam Energy Scan Theory (BEST) Topical Collaboration.

\bibliography{high_baryon_hydro}

\begin{thebibliography}{76}%
\makeatletter
\providecommand \@ifxundefined [1]{%
 \@ifx{#1\undefined}
}%
\providecommand \@ifnum [1]{%
 \ifnum #1\expandafter \@firstoftwo
 \else \expandafter \@secondoftwo
 \fi
}%
\providecommand \@ifx [1]{%
 \ifx #1\expandafter \@firstoftwo
 \else \expandafter \@secondoftwo
 \fi
}%
\providecommand \natexlab [1]{#1}%
\providecommand \enquote  [1]{``#1''}%
\providecommand \bibnamefont  [1]{#1}%
\providecommand \bibfnamefont [1]{#1}%
\providecommand \citenamefont [1]{#1}%
\providecommand \href@noop [0]{\@secondoftwo}%
\providecommand \href [0]{\begingroup \@sanitize@url \@href}%
\providecommand \@href[1]{\@@startlink{#1}\@@href}%
\providecommand \@@href[1]{\endgroup#1\@@endlink}%
\providecommand \@sanitize@url [0]{\catcode `\\12\catcode `\$12\catcode
  `\&12\catcode `\#12\catcode `\^12\catcode `\_12\catcode `\%12\relax}%
\providecommand \@@startlink[1]{}%
\providecommand \@@endlink[0]{}%
\providecommand \url  [0]{\begingroup\@sanitize@url \@url }%
\providecommand \@url [1]{\endgroup\@href {#1}{\urlprefix }}%
\providecommand \urlprefix  [0]{URL }%
\providecommand \Eprint [0]{\href }%
\providecommand \doibase [0]{http://dx.doi.org/}%
\providecommand \selectlanguage [0]{\@gobble}%
\providecommand \bibinfo  [0]{\@secondoftwo}%
\providecommand \bibfield  [0]{\@secondoftwo}%
\providecommand \translation [1]{[#1]}%
\providecommand \BibitemOpen [0]{}%
\providecommand \bibitemStop [0]{}%
\providecommand \bibitemNoStop [0]{.\EOS\space}%
\providecommand \EOS [0]{\spacefactor3000\relax}%
\providecommand \BibitemShut  [1]{\csname bibitem#1\endcsname}%
\let\auto@bib@innerbib\@empty
\bibitem [{\citenamefont {Li}\ and\ \citenamefont
  {Kapusta}(2017)}]{Li:2016wzh}%
  \BibitemOpen
  \bibfield  {author} {\bibinfo {author} {\bibfnamefont {M.}~\bibnamefont
  {Li}}\ and\ \bibinfo {author} {\bibfnamefont {J.~I.}\ \bibnamefont
  {Kapusta}},\ }\href {\doibase 10.1103/PhysRevC.95.011901} {\bibfield
  {journal} {\bibinfo  {journal} {Phys. Rev.}\ }\textbf {\bibinfo {volume}
  {C95}},\ \bibinfo {pages} {011901} (\bibinfo {year} {2017})}\BibitemShut
  {NoStop}%
\bibitem [{\citenamefont {Li}\ and\ \citenamefont
  {Kapusta}(2018)}]{Li:2018ini}%
  \BibitemOpen
  \bibfield  {author} {\bibinfo {author} {\bibfnamefont {M.}~\bibnamefont
  {Li}}\ and\ \bibinfo {author} {\bibfnamefont {J.~I.}\ \bibnamefont
  {Kapusta}},\ }\href@noop {} {\  (\bibinfo {year} {2018})},\ \Eprint
  {http://arxiv.org/abs/1808.05751} {arXiv:1808.05751 [nucl-th]} \BibitemShut
  {NoStop}%
\bibitem [{\citenamefont {Aoki}\ \emph {et~al.}(2006)\citenamefont {Aoki},
  \citenamefont {Endrodi}, \citenamefont {Fodor}, \citenamefont {Katz},\ and\
  \citenamefont {Szabo}}]{Aoki:2006we}%
  \BibitemOpen
  \bibfield  {author} {\bibinfo {author} {\bibfnamefont {Y.}~\bibnamefont
  {Aoki}}, \bibinfo {author} {\bibfnamefont {G.}~\bibnamefont {Endrodi}},
  \bibinfo {author} {\bibfnamefont {Z.}~\bibnamefont {Fodor}}, \bibinfo
  {author} {\bibfnamefont {S.~D.}\ \bibnamefont {Katz}}, \ and\ \bibinfo
  {author} {\bibfnamefont {K.~K.}\ \bibnamefont {Szabo}},\ }\href {\doibase
  10.1038/nature05120} {\bibfield  {journal} {\bibinfo  {journal} {Nature}\
  }\textbf {\bibinfo {volume} {443}},\ \bibinfo {pages} {675} (\bibinfo {year}
  {2006})}\BibitemShut {NoStop}%
\bibitem [{\citenamefont {Stephanov}(2006)}]{Stephanov:2007fk}%
  \BibitemOpen
  \bibfield  {author} {\bibinfo {author} {\bibfnamefont {M.~A.}\ \bibnamefont
  {Stephanov}},\ }\bibfield  {booktitle} {\emph {\bibinfo {booktitle}
  {{Proceedings, 24th International Symposium on Lattice Field Theory (Lattice
  2006): Tucson, USA, July 23-28, 2006}}},\ }\href@noop {} {\bibfield
  {journal} {\bibinfo  {journal} {PoS}\ }\textbf {\bibinfo {volume}
  {LAT2006}},\ \bibinfo {pages} {024} (\bibinfo {year} {2006})}\BibitemShut
  {NoStop}%
\bibitem [{\citenamefont {Adamczyk}\ \emph {et~al.}(2017)\citenamefont
  {Adamczyk} \emph {et~al.}}]{Adamczyk:2017iwn}%
  \BibitemOpen
  \bibfield  {author} {\bibinfo {author} {\bibfnamefont {L.}~\bibnamefont
  {Adamczyk}} \emph {et~al.} (\bibinfo {collaboration} {STAR}),\ }\href
  {\doibase 10.1103/PhysRevC.96.044904} {\bibfield  {journal} {\bibinfo
  {journal} {Phys. Rev.}\ }\textbf {\bibinfo {volume} {C96}},\ \bibinfo {pages}
  {044904} (\bibinfo {year} {2017})},\ \Eprint
  {http://arxiv.org/abs/1701.07065} {arXiv:1701.07065 [nucl-ex]} \BibitemShut
  {NoStop}%
\bibitem [{\citenamefont {Ablyazimov}\ \emph {et~al.}(2017)\citenamefont
  {Ablyazimov} \emph {et~al.}}]{Ablyazimov:2017guv}%
  \BibitemOpen
  \bibfield  {author} {\bibinfo {author} {\bibfnamefont {T.}~\bibnamefont
  {Ablyazimov}} \emph {et~al.} (\bibinfo {collaboration} {CBM}),\ }\href
  {\doibase 10.1140/epja/i2017-12248-y} {\bibfield  {journal} {\bibinfo
  {journal} {Eur. Phys. J.}\ }\textbf {\bibinfo {volume} {A53}},\ \bibinfo
  {pages} {60} (\bibinfo {year} {2017})},\ \Eprint
  {http://arxiv.org/abs/1607.01487} {arXiv:1607.01487 [nucl-ex]} \BibitemShut
  {NoStop}%
\bibitem [{\citenamefont {Sissakian}\ and\ \citenamefont
  {Sorin}(2009)}]{Sissakian:2009zza}%
  \BibitemOpen
  \bibfield  {author} {\bibinfo {author} {\bibfnamefont {A.~N.}\ \bibnamefont
  {Sissakian}}\ and\ \bibinfo {author} {\bibfnamefont {A.~S.}\ \bibnamefont
  {Sorin}} (\bibinfo {collaboration} {NICA}),\ }\bibfield  {booktitle} {\emph
  {\bibinfo {booktitle} {{Strangeness in quark matter. Proceedings,
  International Conference, SQM 2008, Beijing, P.R. China, October 5-10,
  2008}}},\ }\href {\doibase 10.1088/0954-3899/36/6/064069} {\bibfield
  {journal} {\bibinfo  {journal} {J. Phys.}\ }\textbf {\bibinfo {volume}
  {G36}},\ \bibinfo {pages} {064069} (\bibinfo {year} {2009})}\BibitemShut
  {NoStop}%
\bibitem [{\citenamefont {Bearden}\ \emph {et~al.}(2004)\citenamefont {Bearden}
  \emph {et~al.}}]{Bearden:2003hx}%
  \BibitemOpen
  \bibfield  {author} {\bibinfo {author} {\bibfnamefont {I.~G.}\ \bibnamefont
  {Bearden}} \emph {et~al.} (\bibinfo {collaboration} {BRAHMS}),\ }\href
  {\doibase 10.1103/PhysRevLett.93.102301} {\bibfield  {journal} {\bibinfo
  {journal} {Phys. Rev. Lett.}\ }\textbf {\bibinfo {volume} {93}},\ \bibinfo
  {pages} {102301} (\bibinfo {year} {2004})}\BibitemShut {NoStop}%
\bibitem [{\citenamefont {Arsene}\ \emph {et~al.}(2009)\citenamefont {Arsene}
  \emph {et~al.}}]{Arsene:2009aa}%
  \BibitemOpen
  \bibfield  {author} {\bibinfo {author} {\bibfnamefont {I.~C.}\ \bibnamefont
  {Arsene}} \emph {et~al.} (\bibinfo {collaboration} {BRAHMS}),\ }\href
  {\doibase 10.1016/j.physletb.2009.05.049} {\bibfield  {journal} {\bibinfo
  {journal} {Phys. Lett.}\ }\textbf {\bibinfo {volume} {B677}},\ \bibinfo
  {pages} {267} (\bibinfo {year} {2009})}\BibitemShut {NoStop}%
\bibitem [{\citenamefont {Shen}\ and\ \citenamefont
  {Schenke}(2018{\natexlab{a}})}]{Shen:2017bsr}%
  \BibitemOpen
  \bibfield  {author} {\bibinfo {author} {\bibfnamefont {C.}~\bibnamefont
  {Shen}}\ and\ \bibinfo {author} {\bibfnamefont {B.}~\bibnamefont {Schenke}},\
  }\href {\doibase 10.1103/PhysRevC.97.024907} {\bibfield  {journal} {\bibinfo
  {journal} {Phys. Rev.}\ }\textbf {\bibinfo {volume} {C97}},\ \bibinfo {pages}
  {024907} (\bibinfo {year} {2018}{\natexlab{a}})},\ \Eprint
  {http://arxiv.org/abs/1710.00881} {arXiv:1710.00881 [nucl-th]} \BibitemShut
  {NoStop}%
\bibitem [{\citenamefont {Shen}\ and\ \citenamefont
  {Schenke}(2018{\natexlab{b}})}]{Shen:2017fnn}%
  \BibitemOpen
  \bibfield  {author} {\bibinfo {author} {\bibfnamefont {C.}~\bibnamefont
  {Shen}}\ and\ \bibinfo {author} {\bibfnamefont {B.}~\bibnamefont {Schenke}},\
  }\bibfield  {booktitle} {\emph {\bibinfo {booktitle} {{Proceedings, 11th
  International Workshop on Critical Point and Onset of Deconfinement
  (CPOD2017): Stony Brook, NY, USA, August 7-11, 2017}}},\ }\href@noop {}
  {\bibfield  {journal} {\bibinfo  {journal} {PoS}\ }\textbf {\bibinfo {volume}
  {CPOD2017}},\ \bibinfo {pages} {006} (\bibinfo {year}
  {2018}{\natexlab{b}})},\ \Eprint {http://arxiv.org/abs/1711.10544}
  {arXiv:1711.10544 [nucl-th]} \BibitemShut {NoStop}%
\bibitem [{\citenamefont {Bjorken}(1983)}]{Bjorken:1982qr}%
  \BibitemOpen
  \bibfield  {author} {\bibinfo {author} {\bibfnamefont {J.~D.}\ \bibnamefont
  {Bjorken}},\ }\href {\doibase 10.1103/PhysRevD.27.140} {\bibfield  {journal}
  {\bibinfo  {journal} {Phys. Rev.}\ }\textbf {\bibinfo {volume} {D27}},\
  \bibinfo {pages} {140} (\bibinfo {year} {1983})}\BibitemShut {NoStop}%
\bibitem [{\citenamefont {Anishetty}\ \emph {et~al.}(1980)\citenamefont
  {Anishetty}, \citenamefont {Koehler},\ and\ \citenamefont
  {McLerran}}]{Anishetty:1980zp}%
  \BibitemOpen
  \bibfield  {author} {\bibinfo {author} {\bibfnamefont {R.}~\bibnamefont
  {Anishetty}}, \bibinfo {author} {\bibfnamefont {P.}~\bibnamefont {Koehler}},
  \ and\ \bibinfo {author} {\bibfnamefont {L.~D.}\ \bibnamefont {McLerran}},\
  }\href {\doibase 10.1103/PhysRevD.22.2793} {\bibfield  {journal} {\bibinfo
  {journal} {Phys. Rev.}\ }\textbf {\bibinfo {volume} {D22}},\ \bibinfo {pages}
  {2793} (\bibinfo {year} {1980})}\BibitemShut {NoStop}%
\bibitem [{\citenamefont {Gyulassy}\ and\ \citenamefont
  {Csernai}(1986)}]{Gyulassy:1986fk}%
  \BibitemOpen
  \bibfield  {author} {\bibinfo {author} {\bibfnamefont {M.}~\bibnamefont
  {Gyulassy}}\ and\ \bibinfo {author} {\bibfnamefont {L.~P.}\ \bibnamefont
  {Csernai}},\ }\href {\doibase 10.1016/0375-9474(86)90534-8} {\bibfield
  {journal} {\bibinfo  {journal} {Nucl. Phys.}\ }\textbf {\bibinfo {volume}
  {A460}},\ \bibinfo {pages} {723} (\bibinfo {year} {1986})}\BibitemShut
  {NoStop}%
\bibitem [{\citenamefont {McLerran}\ and\ \citenamefont
  {Venugopalan}(1994{\natexlab{a}})}]{McLerran:1993ni}%
  \BibitemOpen
  \bibfield  {author} {\bibinfo {author} {\bibfnamefont {L.~D.}\ \bibnamefont
  {McLerran}}\ and\ \bibinfo {author} {\bibfnamefont {R.}~\bibnamefont
  {Venugopalan}},\ }\href {\doibase 10.1103/PhysRevD.49.2233} {\bibfield
  {journal} {\bibinfo  {journal} {Phys. Rev.}\ }\textbf {\bibinfo {volume}
  {D49}},\ \bibinfo {pages} {2233} (\bibinfo {year}
  {1994}{\natexlab{a}})}\BibitemShut {NoStop}%
\bibitem [{\citenamefont {McLerran}\ and\ \citenamefont
  {Venugopalan}(1994{\natexlab{b}})}]{McLerran:1993ka}%
  \BibitemOpen
  \bibfield  {author} {\bibinfo {author} {\bibfnamefont {L.~D.}\ \bibnamefont
  {McLerran}}\ and\ \bibinfo {author} {\bibfnamefont {R.}~\bibnamefont
  {Venugopalan}},\ }\href {\doibase 10.1103/PhysRevD.49.3352} {\bibfield
  {journal} {\bibinfo  {journal} {Phys. Rev.}\ }\textbf {\bibinfo {volume}
  {D49}},\ \bibinfo {pages} {3352} (\bibinfo {year}
  {1994}{\natexlab{b}})}\BibitemShut {NoStop}%
\bibitem [{\citenamefont {Baier}\ \emph {et~al.}(2001)\citenamefont {Baier},
  \citenamefont {Mueller}, \citenamefont {Schiff},\ and\ \citenamefont
  {Son}}]{Baier:2000sb}%
  \BibitemOpen
  \bibfield  {author} {\bibinfo {author} {\bibfnamefont {R.}~\bibnamefont
  {Baier}}, \bibinfo {author} {\bibfnamefont {A.~H.}\ \bibnamefont {Mueller}},
  \bibinfo {author} {\bibfnamefont {D.}~\bibnamefont {Schiff}}, \ and\ \bibinfo
  {author} {\bibfnamefont {D.~T.}\ \bibnamefont {Son}},\ }\href {\doibase
  10.1016/S0370-2693(01)00191-5} {\bibfield  {journal} {\bibinfo  {journal}
  {Phys. Lett.}\ }\textbf {\bibinfo {volume} {B502}},\ \bibinfo {pages} {51}
  (\bibinfo {year} {2001})}\BibitemShut {NoStop}%
\bibitem [{\citenamefont {Kurkela}\ and\ \citenamefont
  {Zhu}(2015)}]{Kurkela:2015qoa}%
  \BibitemOpen
  \bibfield  {author} {\bibinfo {author} {\bibfnamefont {A.}~\bibnamefont
  {Kurkela}}\ and\ \bibinfo {author} {\bibfnamefont {Y.}~\bibnamefont {Zhu}},\
  }\href {\doibase 10.1103/PhysRevLett.115.182301} {\bibfield  {journal}
  {\bibinfo  {journal} {Phys. Rev. Lett.}\ }\textbf {\bibinfo {volume} {115}},\
  \bibinfo {pages} {182301} (\bibinfo {year} {2015})}\BibitemShut {NoStop}%
\bibitem [{\citenamefont {Romatschke}(2010)}]{Romatschke:2009im}%
  \BibitemOpen
  \bibfield  {author} {\bibinfo {author} {\bibfnamefont {P.}~\bibnamefont
  {Romatschke}},\ }\href {\doibase 10.1142/S0218301310014613} {\bibfield
  {journal} {\bibinfo  {journal} {Int. J. Mod. Phys.}\ }\textbf {\bibinfo
  {volume} {E19}},\ \bibinfo {pages} {1} (\bibinfo {year} {2010})}\BibitemShut
  {NoStop}%
\bibitem [{\citenamefont {Gale}\ \emph {et~al.}(2013)\citenamefont {Gale},
  \citenamefont {Jeon},\ and\ \citenamefont {Schenke}}]{Gale:2013da}%
  \BibitemOpen
  \bibfield  {author} {\bibinfo {author} {\bibfnamefont {C.}~\bibnamefont
  {Gale}}, \bibinfo {author} {\bibfnamefont {S.}~\bibnamefont {Jeon}}, \ and\
  \bibinfo {author} {\bibfnamefont {B.}~\bibnamefont {Schenke}},\ }\href
  {\doibase 10.1142/S0217751X13400113} {\bibfield  {journal} {\bibinfo
  {journal} {Int. J. Mod. Phys.}\ }\textbf {\bibinfo {volume} {A28}},\ \bibinfo
  {pages} {1340011} (\bibinfo {year} {2013})}\BibitemShut {NoStop}%
\bibitem [{\citenamefont {Jeon}\ and\ \citenamefont
  {Heinz}(2015)}]{Jeon:2015dfa}%
  \BibitemOpen
  \bibfield  {author} {\bibinfo {author} {\bibfnamefont {S.}~\bibnamefont
  {Jeon}}\ and\ \bibinfo {author} {\bibfnamefont {U.}~\bibnamefont {Heinz}},\
  }\href {\doibase 10.1142/S0218301315300106} {\bibfield  {journal} {\bibinfo
  {journal} {Int. J. Mod. Phys.}\ }\textbf {\bibinfo {volume} {E24}},\ \bibinfo
  {pages} {1530010} (\bibinfo {year} {2015})}\BibitemShut {NoStop}%
\bibitem [{\citenamefont {Jaiswal}\ and\ \citenamefont
  {Roy}(2016)}]{Jaiswal:2016hex}%
  \BibitemOpen
  \bibfield  {author} {\bibinfo {author} {\bibfnamefont {A.}~\bibnamefont
  {Jaiswal}}\ and\ \bibinfo {author} {\bibfnamefont {V.}~\bibnamefont {Roy}},\
  }\href {\doibase 10.1155/2016/9623034} {\bibfield  {journal} {\bibinfo
  {journal} {Adv. High Energy Phys.}\ }\textbf {\bibinfo {volume} {2016}},\
  \bibinfo {pages} {9623034} (\bibinfo {year} {2016})}\BibitemShut {NoStop}%
\bibitem [{\citenamefont {Romatschke}\ and\ \citenamefont
  {Romatschke}(2017)}]{Romatschke:2017ejr}%
  \BibitemOpen
  \bibfield  {author} {\bibinfo {author} {\bibfnamefont {P.}~\bibnamefont
  {Romatschke}}\ and\ \bibinfo {author} {\bibfnamefont {U.}~\bibnamefont
  {Romatschke}},\ }\href@noop {} {\  (\bibinfo {year} {2017})},\ \Eprint
  {http://arxiv.org/abs/1712.05815} {arXiv:1712.05815 [nucl-th]} \BibitemShut
  {NoStop}%
\bibitem [{\citenamefont {Denicol}\ \emph {et~al.}(2012)\citenamefont
  {Denicol}, \citenamefont {Niemi}, \citenamefont {Molnar},\ and\ \citenamefont
  {Rischke}}]{Denicol:2012cn}%
  \BibitemOpen
  \bibfield  {author} {\bibinfo {author} {\bibfnamefont {G.~S.}\ \bibnamefont
  {Denicol}}, \bibinfo {author} {\bibfnamefont {H.}~\bibnamefont {Niemi}},
  \bibinfo {author} {\bibfnamefont {E.}~\bibnamefont {Molnar}}, \ and\ \bibinfo
  {author} {\bibfnamefont {D.~H.}\ \bibnamefont {Rischke}},\ }\href {\doibase
  10.1103/PhysRevD.85.114047, 10.1103/PhysRevD.91.039902} {\bibfield  {journal}
  {\bibinfo  {journal} {Phys. Rev.}\ }\textbf {\bibinfo {volume} {D85}},\
  \bibinfo {pages} {114047} (\bibinfo {year} {2012})},\ \bibinfo {note}
  {[Erratum: Phys. Rev.D91,no.3,039902(2015)]}\BibitemShut {NoStop}%
\bibitem [{\citenamefont {Israel}\ and\ \citenamefont
  {Stewart}(1979)}]{Israel:1979wp}%
  \BibitemOpen
  \bibfield  {author} {\bibinfo {author} {\bibfnamefont {W.}~\bibnamefont
  {Israel}}\ and\ \bibinfo {author} {\bibfnamefont {J.~M.}\ \bibnamefont
  {Stewart}},\ }\href {\doibase 10.1016/0003-4916(79)90130-1} {\bibfield
  {journal} {\bibinfo  {journal} {Annals Phys.}\ }\textbf {\bibinfo {volume}
  {118}},\ \bibinfo {pages} {341} (\bibinfo {year} {1979})}\BibitemShut
  {NoStop}%
\bibitem [{\citenamefont {Baier}\ \emph {et~al.}(2008)\citenamefont {Baier},
  \citenamefont {Romatschke}, \citenamefont {Son}, \citenamefont {Starinets},\
  and\ \citenamefont {Stephanov}}]{Baier:2007ix}%
  \BibitemOpen
  \bibfield  {author} {\bibinfo {author} {\bibfnamefont {R.}~\bibnamefont
  {Baier}}, \bibinfo {author} {\bibfnamefont {P.}~\bibnamefont {Romatschke}},
  \bibinfo {author} {\bibfnamefont {D.~T.}\ \bibnamefont {Son}}, \bibinfo
  {author} {\bibfnamefont {A.~O.}\ \bibnamefont {Starinets}}, \ and\ \bibinfo
  {author} {\bibfnamefont {M.~A.}\ \bibnamefont {Stephanov}},\ }\href {\doibase
  10.1088/1126-6708/2008/04/100} {\bibfield  {journal} {\bibinfo  {journal}
  {JHEP}\ }\textbf {\bibinfo {volume} {04}},\ \bibinfo {pages} {100} (\bibinfo
  {year} {2008})}\BibitemShut {NoStop}%
\bibitem [{\citenamefont {Song}(2013)}]{Song:2012ua}%
  \BibitemOpen
  \bibfield  {author} {\bibinfo {author} {\bibfnamefont {H.}~\bibnamefont
  {Song}},\ }\href {\doibase 10.1016/j.nuclphysa.2013.01.052} {\bibfield
  {journal} {\bibinfo  {journal} {Nucl. Phys.}\ }\textbf {\bibinfo {volume}
  {A904-905}},\ \bibinfo {pages} {114c} (\bibinfo {year} {2013})}\BibitemShut
  {NoStop}%
\bibitem [{\citenamefont {Denicol}\ \emph {et~al.}(2018)\citenamefont
  {Denicol}, \citenamefont {Gale}, \citenamefont {Jeon}, \citenamefont
  {Monnai}, \citenamefont {Schenke},\ and\ \citenamefont
  {Shen}}]{Denicol:2018wdp}%
  \BibitemOpen
  \bibfield  {author} {\bibinfo {author} {\bibfnamefont {G.~S.}\ \bibnamefont
  {Denicol}}, \bibinfo {author} {\bibfnamefont {C.}~\bibnamefont {Gale}},
  \bibinfo {author} {\bibfnamefont {S.}~\bibnamefont {Jeon}}, \bibinfo {author}
  {\bibfnamefont {A.}~\bibnamefont {Monnai}}, \bibinfo {author} {\bibfnamefont
  {B.}~\bibnamefont {Schenke}}, \ and\ \bibinfo {author} {\bibfnamefont
  {C.}~\bibnamefont {Shen}},\ }\href@noop {} {\  (\bibinfo {year} {2018})},\
  \Eprint {http://arxiv.org/abs/1804.10557} {arXiv:1804.10557 [nucl-th]}
  \BibitemShut {NoStop}%
\bibitem [{\citenamefont {Satarov}\ \emph {et~al.}(2007)\citenamefont
  {Satarov}, \citenamefont {Merdeev}, \citenamefont {Mishustin},\ and\
  \citenamefont {Stoecker}}]{Satarov:2006iw}%
  \BibitemOpen
  \bibfield  {author} {\bibinfo {author} {\bibfnamefont {L.~M.}\ \bibnamefont
  {Satarov}}, \bibinfo {author} {\bibfnamefont {A.~V.}\ \bibnamefont
  {Merdeev}}, \bibinfo {author} {\bibfnamefont {I.~N.}\ \bibnamefont
  {Mishustin}}, \ and\ \bibinfo {author} {\bibfnamefont {H.}~\bibnamefont
  {Stoecker}},\ }\href {\doibase 10.1103/PhysRevC.75.024903} {\bibfield
  {journal} {\bibinfo  {journal} {Phys. Rev.}\ }\textbf {\bibinfo {volume}
  {C75}},\ \bibinfo {pages} {024903} (\bibinfo {year} {2007})}\BibitemShut
  {NoStop}%
\bibitem [{\citenamefont {Bozek}(2008)}]{Bozek:2007qt}%
  \BibitemOpen
  \bibfield  {author} {\bibinfo {author} {\bibfnamefont {P.}~\bibnamefont
  {Bozek}},\ }\href {\doibase 10.1103/PhysRevC.77.034911} {\bibfield  {journal}
  {\bibinfo  {journal} {Phys. Rev.}\ }\textbf {\bibinfo {volume} {C77}},\
  \bibinfo {pages} {034911} (\bibinfo {year} {2008})}\BibitemShut {NoStop}%
\bibitem [{\citenamefont {Monnai}(2012)}]{Monnai:2012jc}%
  \BibitemOpen
  \bibfield  {author} {\bibinfo {author} {\bibfnamefont {A.}~\bibnamefont
  {Monnai}},\ }\href {\doibase 10.1103/PhysRevC.86.014908} {\bibfield
  {journal} {\bibinfo  {journal} {Phys. Rev.}\ }\textbf {\bibinfo {volume}
  {C86}},\ \bibinfo {pages} {014908} (\bibinfo {year} {2012})}\BibitemShut
  {NoStop}%
\bibitem [{\citenamefont {Florkowski}\ \emph {et~al.}(2016)\citenamefont
  {Florkowski}, \citenamefont {Ryblewski}, \citenamefont {Strickland},\ and\
  \citenamefont {Tinti}}]{Florkowski:2016kjj}%
  \BibitemOpen
  \bibfield  {author} {\bibinfo {author} {\bibfnamefont {W.}~\bibnamefont
  {Florkowski}}, \bibinfo {author} {\bibfnamefont {R.}~\bibnamefont
  {Ryblewski}}, \bibinfo {author} {\bibfnamefont {M.}~\bibnamefont
  {Strickland}}, \ and\ \bibinfo {author} {\bibfnamefont {L.}~\bibnamefont
  {Tinti}},\ }\href {\doibase 10.1103/PhysRevC.94.064903} {\bibfield  {journal}
  {\bibinfo  {journal} {Phys. Rev.}\ }\textbf {\bibinfo {volume} {C94}},\
  \bibinfo {pages} {064903} (\bibinfo {year} {2016})}\BibitemShut {NoStop}%
\bibitem [{\citenamefont {Cattaneo}(1958)}]{Cattaneo:1958}%
  \BibitemOpen
  \bibfield  {author} {\bibinfo {author} {\bibfnamefont {C.~R.}\ \bibnamefont
  {Cattaneo}},\ }\href@noop {} {\bibfield  {journal} {\bibinfo  {journal}
  {Comptes Rendus.}\ }\textbf {\bibinfo {volume} {247}},\ \bibinfo {pages}
  {431} (\bibinfo {year} {1958})}\BibitemShut {NoStop}%
\bibitem [{\citenamefont {Kapusta}\ and\ \citenamefont
  {Plumberg}(2018)}]{Kapusta:2017hfi}%
  \BibitemOpen
  \bibfield  {author} {\bibinfo {author} {\bibfnamefont {J.~I.}\ \bibnamefont
  {Kapusta}}\ and\ \bibinfo {author} {\bibfnamefont {C.}~\bibnamefont
  {Plumberg}},\ }\href {\doibase 10.1103/PhysRevC.97.014906} {\bibfield
  {journal} {\bibinfo  {journal} {Phys. Rev.}\ }\textbf {\bibinfo {volume}
  {C97}},\ \bibinfo {pages} {014906} (\bibinfo {year} {2018})}\BibitemShut
  {NoStop}%
\bibitem [{\citenamefont {Kapusta}\ and\ \citenamefont
  {Young}(2014)}]{Kapusta:2014dja}%
  \BibitemOpen
  \bibfield  {author} {\bibinfo {author} {\bibfnamefont {J.~I.}\ \bibnamefont
  {Kapusta}}\ and\ \bibinfo {author} {\bibfnamefont {C.}~\bibnamefont
  {Young}},\ }\href {\doibase 10.1103/PhysRevC.90.044902} {\bibfield  {journal}
  {\bibinfo  {journal} {Phys. Rev.}\ }\textbf {\bibinfo {volume} {C90}},\
  \bibinfo {pages} {044902} (\bibinfo {year} {2014})}\BibitemShut {NoStop}%
\bibitem [{\citenamefont {Kurganov}(2016)}]{Kurganov:2016}%
  \BibitemOpen
  \bibfield  {author} {\bibinfo {author} {\bibfnamefont {A.}~\bibnamefont
  {Kurganov}},\ }in\ \href@noop {} {\emph {\bibinfo {booktitle} {Handbook of
  Numerical Analysis}}},\ \bibinfo {editor} {edited by\ \bibinfo {editor}
  {\bibfnamefont {R.}~\bibnamefont {Abgrall}}\ and\ \bibinfo {editor}
  {\bibfnamefont {C.-W.}\ \bibnamefont {Shu}}}\ (\bibinfo  {publisher}
  {Elsevier},\ \bibinfo {year} {2016})\ Chap.~\bibinfo {chapter} {20}, pp.\
  \bibinfo {pages} {525--584}\BibitemShut {NoStop}%
\bibitem [{\citenamefont {Kurganov}\ and\ \citenamefont
  {Lin}(2007)}]{KurganovLin:2007}%
  \BibitemOpen
  \bibfield  {author} {\bibinfo {author} {\bibfnamefont {A.}~\bibnamefont
  {Kurganov}}\ and\ \bibinfo {author} {\bibfnamefont {C.-T.}\ \bibnamefont
  {Lin}},\ }\href@noop {} {\bibfield  {journal} {\bibinfo  {journal} {Commun.
  Comput. Phys.}\ }\textbf {\bibinfo {volume} {2}},\ \bibinfo {pages} {141}
  (\bibinfo {year} {2007})}\BibitemShut {NoStop}%
\bibitem [{\citenamefont {Kurganov}\ and\ \citenamefont
  {Tadmor}(2000)}]{KurganovTadmor:2000}%
  \BibitemOpen
  \bibfield  {author} {\bibinfo {author} {\bibfnamefont {A.}~\bibnamefont
  {Kurganov}}\ and\ \bibinfo {author} {\bibfnamefont {E.}~\bibnamefont
  {Tadmor}},\ }\href@noop {} {\bibfield  {journal} {\bibinfo  {journal} {J.
  Comput. Phys.}\ }\textbf {\bibinfo {volume} {160}},\ \bibinfo {pages} {241}
  (\bibinfo {year} {2000})}\BibitemShut {NoStop}%
\bibitem [{\citenamefont {Schenke}\ \emph {et~al.}(2010)\citenamefont
  {Schenke}, \citenamefont {Jeon},\ and\ \citenamefont
  {Gale}}]{Schenke:2010nt}%
  \BibitemOpen
  \bibfield  {author} {\bibinfo {author} {\bibfnamefont {B.}~\bibnamefont
  {Schenke}}, \bibinfo {author} {\bibfnamefont {S.}~\bibnamefont {Jeon}}, \
  and\ \bibinfo {author} {\bibfnamefont {C.}~\bibnamefont {Gale}},\ }\href
  {\doibase 10.1103/PhysRevC.82.014903} {\bibfield  {journal} {\bibinfo
  {journal} {Phys. Rev.}\ }\textbf {\bibinfo {volume} {C82}},\ \bibinfo {pages}
  {014903} (\bibinfo {year} {2010})}\BibitemShut {NoStop}%
\bibitem [{\citenamefont {Schenke}\ \emph {et~al.}(2011)\citenamefont
  {Schenke}, \citenamefont {Jeon},\ and\ \citenamefont
  {Gale}}]{Schenke:2010rr}%
  \BibitemOpen
  \bibfield  {author} {\bibinfo {author} {\bibfnamefont {B.}~\bibnamefont
  {Schenke}}, \bibinfo {author} {\bibfnamefont {S.}~\bibnamefont {Jeon}}, \
  and\ \bibinfo {author} {\bibfnamefont {C.}~\bibnamefont {Gale}},\ }\href
  {\doibase 10.1103/PhysRevLett.106.042301} {\bibfield  {journal} {\bibinfo
  {journal} {Phys. Rev. Lett.}\ }\textbf {\bibinfo {volume} {106}},\ \bibinfo
  {pages} {042301} (\bibinfo {year} {2011})}\BibitemShut {NoStop}%
\bibitem [{\citenamefont {Paquet}\ \emph {et~al.}(2016)\citenamefont {Paquet},
  \citenamefont {Shen}, \citenamefont {Denicol}, \citenamefont {Luzum},
  \citenamefont {Schenke}, \citenamefont {Jeon},\ and\ \citenamefont
  {Gale}}]{Paquet:2015lta}%
  \BibitemOpen
  \bibfield  {author} {\bibinfo {author} {\bibfnamefont {J.-F.}\ \bibnamefont
  {Paquet}}, \bibinfo {author} {\bibfnamefont {C.}~\bibnamefont {Shen}},
  \bibinfo {author} {\bibfnamefont {G.~S.}\ \bibnamefont {Denicol}}, \bibinfo
  {author} {\bibfnamefont {M.}~\bibnamefont {Luzum}}, \bibinfo {author}
  {\bibfnamefont {B.}~\bibnamefont {Schenke}}, \bibinfo {author} {\bibfnamefont
  {S.}~\bibnamefont {Jeon}}, \ and\ \bibinfo {author} {\bibfnamefont
  {C.}~\bibnamefont {Gale}},\ }\href {\doibase 10.1103/PhysRevC.93.044906}
  {\bibfield  {journal} {\bibinfo  {journal} {Phys. Rev.}\ }\textbf {\bibinfo
  {volume} {C93}},\ \bibinfo {pages} {044906} (\bibinfo {year}
  {2016})}\BibitemShut {NoStop}%
\bibitem [{\citenamefont {Song}\ \emph {et~al.}(2011)\citenamefont {Song},
  \citenamefont {Bass},\ and\ \citenamefont {Heinz}}]{Song:2010aq}%
  \BibitemOpen
  \bibfield  {author} {\bibinfo {author} {\bibfnamefont {H.}~\bibnamefont
  {Song}}, \bibinfo {author} {\bibfnamefont {S.~A.}\ \bibnamefont {Bass}}, \
  and\ \bibinfo {author} {\bibfnamefont {U.}~\bibnamefont {Heinz}},\ }\href
  {\doibase 10.1103/PhysRevC.83.024912} {\bibfield  {journal} {\bibinfo
  {journal} {Phys. Rev.}\ }\textbf {\bibinfo {volume} {C83}},\ \bibinfo {pages}
  {024912} (\bibinfo {year} {2011})}\BibitemShut {NoStop}%
\bibitem [{\citenamefont {Hirano}(2002)}]{Hirano:2001eu}%
  \BibitemOpen
  \bibfield  {author} {\bibinfo {author} {\bibfnamefont {T.}~\bibnamefont
  {Hirano}},\ }\href {\doibase 10.1103/PhysRevC.65.011901} {\bibfield
  {journal} {\bibinfo  {journal} {Phys. Rev.}\ }\textbf {\bibinfo {volume}
  {C65}},\ \bibinfo {pages} {011901} (\bibinfo {year} {2002})}\BibitemShut
  {NoStop}%
\bibitem [{\citenamefont {Hirano}\ and\ \citenamefont
  {Tsuda}(2002)}]{Hirano:2002ds}%
  \BibitemOpen
  \bibfield  {author} {\bibinfo {author} {\bibfnamefont {T.}~\bibnamefont
  {Hirano}}\ and\ \bibinfo {author} {\bibfnamefont {K.}~\bibnamefont {Tsuda}},\
  }\href {\doibase 10.1103/PhysRevC.66.054905} {\bibfield  {journal} {\bibinfo
  {journal} {Phys. Rev.}\ }\textbf {\bibinfo {volume} {C66}},\ \bibinfo {pages}
  {054905} (\bibinfo {year} {2002})}\BibitemShut {NoStop}%
\bibitem [{\citenamefont {Parotto}\ \emph {et~al.}(2018)\citenamefont
  {Parotto}, \citenamefont {Bluhm}, \citenamefont {Mroczek}, \citenamefont
  {Nahrgang}, \citenamefont {Noronha-Hostler}, \citenamefont {Rajagopal},
  \citenamefont {Ratti}, \citenamefont {Schäfer},\ and\ \citenamefont
  {Stephanov}}]{Parotto:2018pwx}%
  \BibitemOpen
  \bibfield  {author} {\bibinfo {author} {\bibfnamefont {P.}~\bibnamefont
  {Parotto}}, \bibinfo {author} {\bibfnamefont {M.}~\bibnamefont {Bluhm}},
  \bibinfo {author} {\bibfnamefont {D.}~\bibnamefont {Mroczek}}, \bibinfo
  {author} {\bibfnamefont {M.}~\bibnamefont {Nahrgang}}, \bibinfo {author}
  {\bibfnamefont {J.}~\bibnamefont {Noronha-Hostler}}, \bibinfo {author}
  {\bibfnamefont {K.}~\bibnamefont {Rajagopal}}, \bibinfo {author}
  {\bibfnamefont {C.}~\bibnamefont {Ratti}}, \bibinfo {author} {\bibfnamefont
  {T.}~\bibnamefont {Schäfer}}, \ and\ \bibinfo {author} {\bibfnamefont
  {M.}~\bibnamefont {Stephanov}},\ }\href@noop {} {\  (\bibinfo {year}
  {2018})}\BibitemShut {NoStop}%
\bibitem [{\citenamefont {Albright}\ \emph {et~al.}(2014)\citenamefont
  {Albright}, \citenamefont {Kapusta},\ and\ \citenamefont
  {Young}}]{Albright:2014gva}%
  \BibitemOpen
  \bibfield  {author} {\bibinfo {author} {\bibfnamefont {M.}~\bibnamefont
  {Albright}}, \bibinfo {author} {\bibfnamefont {J.}~\bibnamefont {Kapusta}}, \
  and\ \bibinfo {author} {\bibfnamefont {C.}~\bibnamefont {Young}},\ }\href
  {\doibase 10.1103/PhysRevC.90.024915} {\bibfield  {journal} {\bibinfo
  {journal} {Phys. Rev.}\ }\textbf {\bibinfo {volume} {C90}},\ \bibinfo {pages}
  {024915} (\bibinfo {year} {2014})}\BibitemShut {NoStop}%
\bibitem [{\citenamefont {Albright}\ \emph {et~al.}(2015)\citenamefont
  {Albright}, \citenamefont {Kapusta},\ and\ \citenamefont
  {Young}}]{Albright:2015uua}%
  \BibitemOpen
  \bibfield  {author} {\bibinfo {author} {\bibfnamefont {M.}~\bibnamefont
  {Albright}}, \bibinfo {author} {\bibfnamefont {J.}~\bibnamefont {Kapusta}}, \
  and\ \bibinfo {author} {\bibfnamefont {C.}~\bibnamefont {Young}},\ }\href
  {\doibase 10.1103/PhysRevC.92.044904} {\bibfield  {journal} {\bibinfo
  {journal} {Phys. Rev.}\ }\textbf {\bibinfo {volume} {C92}},\ \bibinfo {pages}
  {044904} (\bibinfo {year} {2015})}\BibitemShut {NoStop}%
\bibitem [{\citenamefont {Albright}\ and\ \citenamefont
  {Kapusta}(2016)}]{Albright:2015fpa}%
  \BibitemOpen
  \bibfield  {author} {\bibinfo {author} {\bibfnamefont {M.}~\bibnamefont
  {Albright}}\ and\ \bibinfo {author} {\bibfnamefont {J.~I.}\ \bibnamefont
  {Kapusta}},\ }\href {\doibase 10.1103/PhysRevC.93.014903} {\bibfield
  {journal} {\bibinfo  {journal} {Phys. Rev.}\ }\textbf {\bibinfo {volume}
  {C93}},\ \bibinfo {pages} {014903} (\bibinfo {year} {2016})}\BibitemShut
  {NoStop}%
\bibitem [{\citenamefont {Jaiswal}\ \emph {et~al.}(2015)\citenamefont
  {Jaiswal}, \citenamefont {Friman},\ and\ \citenamefont
  {Redlich}}]{Jaiswal:2015mxa}%
  \BibitemOpen
  \bibfield  {author} {\bibinfo {author} {\bibfnamefont {A.}~\bibnamefont
  {Jaiswal}}, \bibinfo {author} {\bibfnamefont {B.}~\bibnamefont {Friman}}, \
  and\ \bibinfo {author} {\bibfnamefont {K.}~\bibnamefont {Redlich}},\ }\href
  {\doibase 10.1016/j.physletb.2015.11.018} {\bibfield  {journal} {\bibinfo
  {journal} {Phys. Lett.}\ }\textbf {\bibinfo {volume} {B751}},\ \bibinfo
  {pages} {548} (\bibinfo {year} {2015})}\BibitemShut {NoStop}%
\bibitem [{\citenamefont {Greif}\ \emph {et~al.}(2018)\citenamefont {Greif},
  \citenamefont {Fotakis}, \citenamefont {Denicol},\ and\ \citenamefont
  {Greiner}}]{Greif:2017byw}%
  \BibitemOpen
  \bibfield  {author} {\bibinfo {author} {\bibfnamefont {M.}~\bibnamefont
  {Greif}}, \bibinfo {author} {\bibfnamefont {J.~A.}\ \bibnamefont {Fotakis}},
  \bibinfo {author} {\bibfnamefont {G.~S.}\ \bibnamefont {Denicol}}, \ and\
  \bibinfo {author} {\bibfnamefont {C.}~\bibnamefont {Greiner}},\ }\href
  {\doibase 10.1103/PhysRevLett.120.242301} {\bibfield  {journal} {\bibinfo
  {journal} {Phys. Rev. Lett.}\ }\textbf {\bibinfo {volume} {120}},\ \bibinfo
  {pages} {242301} (\bibinfo {year} {2018})}\BibitemShut {NoStop}%
\bibitem [{\citenamefont {Son}\ and\ \citenamefont
  {Starinets}(2006)}]{Son:2006em}%
  \BibitemOpen
  \bibfield  {author} {\bibinfo {author} {\bibfnamefont {D.~T.}\ \bibnamefont
  {Son}}\ and\ \bibinfo {author} {\bibfnamefont {A.~O.}\ \bibnamefont
  {Starinets}},\ }\href {\doibase 10.1088/1126-6708/2006/03/052} {\bibfield
  {journal} {\bibinfo  {journal} {JHEP}\ }\textbf {\bibinfo {volume} {03}},\
  \bibinfo {pages} {052} (\bibinfo {year} {2006})}\BibitemShut {NoStop}%
\bibitem [{\citenamefont {Rougemont}\ \emph {et~al.}(2015)\citenamefont
  {Rougemont}, \citenamefont {Noronha},\ and\ \citenamefont
  {Noronha-Hostler}}]{Rougemont:2015ona}%
  \BibitemOpen
  \bibfield  {author} {\bibinfo {author} {\bibfnamefont {R.}~\bibnamefont
  {Rougemont}}, \bibinfo {author} {\bibfnamefont {J.}~\bibnamefont {Noronha}},
  \ and\ \bibinfo {author} {\bibfnamefont {J.}~\bibnamefont
  {Noronha-Hostler}},\ }\href {\doibase 10.1103/PhysRevLett.115.202301}
  {\bibfield  {journal} {\bibinfo  {journal} {Phys. Rev. Lett.}\ }\textbf
  {\bibinfo {volume} {115}},\ \bibinfo {pages} {202301} (\bibinfo {year}
  {2015})}\BibitemShut {NoStop}%
\bibitem [{\citenamefont {Danielewicz}\ and\ \citenamefont
  {Gyulassy}(1985)}]{Danielewicz:1984ww}%
  \BibitemOpen
  \bibfield  {author} {\bibinfo {author} {\bibfnamefont {P.}~\bibnamefont
  {Danielewicz}}\ and\ \bibinfo {author} {\bibfnamefont {M.}~\bibnamefont
  {Gyulassy}},\ }\href {\doibase 10.1103/PhysRevD.31.53} {\bibfield  {journal}
  {\bibinfo  {journal} {Phys. Rev.}\ }\textbf {\bibinfo {volume} {D31}},\
  \bibinfo {pages} {53} (\bibinfo {year} {1985})}\BibitemShut {NoStop}%
\bibitem [{\citenamefont {Hosoya}\ and\ \citenamefont
  {Kajantie}(1985)}]{Hosoya:1983xm}%
  \BibitemOpen
  \bibfield  {author} {\bibinfo {author} {\bibfnamefont {A.}~\bibnamefont
  {Hosoya}}\ and\ \bibinfo {author} {\bibfnamefont {K.}~\bibnamefont
  {Kajantie}},\ }\href {\doibase 10.1016/0550-3213(85)90499-7} {\bibfield
  {journal} {\bibinfo  {journal} {Nucl. Phys.}\ }\textbf {\bibinfo {volume}
  {B250}},\ \bibinfo {pages} {666} (\bibinfo {year} {1985})}\BibitemShut
  {NoStop}%
\bibitem [{\citenamefont {Gavin}(1985)}]{Gavin:1985ph}%
  \BibitemOpen
  \bibfield  {author} {\bibinfo {author} {\bibfnamefont {S.}~\bibnamefont
  {Gavin}},\ }\href {\doibase 10.1016/0375-9474(85)90190-3} {\bibfield
  {journal} {\bibinfo  {journal} {Nucl. Phys.}\ }\textbf {\bibinfo {volume}
  {A435}},\ \bibinfo {pages} {826} (\bibinfo {year} {1985})}\BibitemShut
  {NoStop}%
\bibitem [{\citenamefont {Heiselberg}\ and\ \citenamefont
  {Pethick}(1993)}]{Heiselberg:1993cr}%
  \BibitemOpen
  \bibfield  {author} {\bibinfo {author} {\bibfnamefont {H.}~\bibnamefont
  {Heiselberg}}\ and\ \bibinfo {author} {\bibfnamefont {C.~J.}\ \bibnamefont
  {Pethick}},\ }\href {\doibase 10.1103/PhysRevD.48.2916} {\bibfield  {journal}
  {\bibinfo  {journal} {Phys. Rev.}\ }\textbf {\bibinfo {volume} {D48}},\
  \bibinfo {pages} {2916} (\bibinfo {year} {1993})}\BibitemShut {NoStop}%
\bibitem [{\citenamefont {Floerchinger}\ and\ \citenamefont
  {Martinez}(2015)}]{Floerchinger:2015efa}%
  \BibitemOpen
  \bibfield  {author} {\bibinfo {author} {\bibfnamefont {S.}~\bibnamefont
  {Floerchinger}}\ and\ \bibinfo {author} {\bibfnamefont {M.}~\bibnamefont
  {Martinez}},\ }\href {\doibase 10.1103/PhysRevC.92.064906} {\bibfield
  {journal} {\bibinfo  {journal} {Phys. Rev.}\ }\textbf {\bibinfo {volume}
  {C92}},\ \bibinfo {pages} {064906} (\bibinfo {year} {2015})}\BibitemShut
  {NoStop}%
\bibitem [{\citenamefont {Bass}\ \emph {et~al.}(1998)\citenamefont {Bass} \emph
  {et~al.}}]{Bass:1998ca}%
  \BibitemOpen
  \bibfield  {author} {\bibinfo {author} {\bibfnamefont {S.~A.}\ \bibnamefont
  {Bass}} \emph {et~al.},\ }\href {\doibase 10.1016/S0146-6410(98)00058-1}
  {\bibfield  {journal} {\bibinfo  {journal} {Prog. Part. Nucl. Phys.}\
  }\textbf {\bibinfo {volume} {41}},\ \bibinfo {pages} {255} (\bibinfo {year}
  {1998})}\BibitemShut {NoStop}%
\bibitem [{\citenamefont {Bleicher}\ \emph {et~al.}(1999)\citenamefont
  {Bleicher} \emph {et~al.}}]{Bleicher:1999xi}%
  \BibitemOpen
  \bibfield  {author} {\bibinfo {author} {\bibfnamefont {M.}~\bibnamefont
  {Bleicher}} \emph {et~al.},\ }\href {\doibase 10.1088/0954-3899/25/9/308}
  {\bibfield  {journal} {\bibinfo  {journal} {J. Phys.}\ }\textbf {\bibinfo
  {volume} {G25}},\ \bibinfo {pages} {1859} (\bibinfo {year}
  {1999})}\BibitemShut {NoStop}%
\bibitem [{\citenamefont {Cooper}\ and\ \citenamefont
  {Frye}(1974)}]{Cooper:1974mv}%
  \BibitemOpen
  \bibfield  {author} {\bibinfo {author} {\bibfnamefont {F.}~\bibnamefont
  {Cooper}}\ and\ \bibinfo {author} {\bibfnamefont {G.}~\bibnamefont {Frye}},\
  }\href {\doibase 10.1103/PhysRevD.10.186} {\bibfield  {journal} {\bibinfo
  {journal} {Phys. Rev.}\ }\textbf {\bibinfo {volume} {D10}},\ \bibinfo {pages}
  {186} (\bibinfo {year} {1974})}\BibitemShut {NoStop}%
\bibitem [{\citenamefont {Kapusta}(1977)}]{Kapusta:1977ce}%
  \BibitemOpen
  \bibfield  {author} {\bibinfo {author} {\bibfnamefont {J.~I.}\ \bibnamefont
  {Kapusta}},\ }\href {\doibase 10.1103/PhysRevC.16.1493} {\bibfield  {journal}
  {\bibinfo  {journal} {Phys. Rev.}\ }\textbf {\bibinfo {volume} {C16}},\
  \bibinfo {pages} {1493} (\bibinfo {year} {1977})}\BibitemShut {NoStop}%
\bibitem [{\citenamefont {Sollfrank}\ \emph {et~al.}(1990)\citenamefont
  {Sollfrank}, \citenamefont {Koch},\ and\ \citenamefont
  {Heinz}}]{Sollfrank:1990qz}%
  \BibitemOpen
  \bibfield  {author} {\bibinfo {author} {\bibfnamefont {J.}~\bibnamefont
  {Sollfrank}}, \bibinfo {author} {\bibfnamefont {P.}~\bibnamefont {Koch}}, \
  and\ \bibinfo {author} {\bibfnamefont {U.~W.}\ \bibnamefont {Heinz}},\ }\href
  {\doibase 10.1016/0370-2693(90)90870-C} {\bibfield  {journal} {\bibinfo
  {journal} {Phys. Lett.}\ }\textbf {\bibinfo {volume} {B252}},\ \bibinfo
  {pages} {256} (\bibinfo {year} {1990})}\BibitemShut {NoStop}%
\bibitem [{\citenamefont {Sollfrank}\ \emph {et~al.}(1991)\citenamefont
  {Sollfrank}, \citenamefont {Koch},\ and\ \citenamefont
  {Heinz}}]{Sollfrank:1991xm}%
  \BibitemOpen
  \bibfield  {author} {\bibinfo {author} {\bibfnamefont {J.}~\bibnamefont
  {Sollfrank}}, \bibinfo {author} {\bibfnamefont {P.}~\bibnamefont {Koch}}, \
  and\ \bibinfo {author} {\bibfnamefont {U.~W.}\ \bibnamefont {Heinz}},\ }\href
  {\doibase 10.1007/BF01562334} {\bibfield  {journal} {\bibinfo  {journal} {Z.
  Phys.}\ }\textbf {\bibinfo {volume} {C52}},\ \bibinfo {pages} {593} (\bibinfo
  {year} {1991})}\BibitemShut {NoStop}%
\bibitem [{\citenamefont {Gorenstein}\ \emph {et~al.}(1995)\citenamefont
  {Gorenstein}, \citenamefont {Tsai},\ and\ \citenamefont
  {Yang}}]{Gorenstein:1987zm}%
  \BibitemOpen
  \bibfield  {author} {\bibinfo {author} {\bibfnamefont {M.~I.}\ \bibnamefont
  {Gorenstein}}, \bibinfo {author} {\bibfnamefont {M.-S.}\ \bibnamefont
  {Tsai}}, \ and\ \bibinfo {author} {\bibfnamefont {S.-N.}\ \bibnamefont
  {Yang}},\ }\href {\doibase 10.1103/PhysRevC.51.1465} {\bibfield  {journal}
  {\bibinfo  {journal} {Phys. Rev.}\ }\textbf {\bibinfo {volume} {C51}},\
  \bibinfo {pages} {1465} (\bibinfo {year} {1995})}\BibitemShut {NoStop}%
\bibitem [{\citenamefont {Lo}(2018)}]{Lo:2017sux}%
  \BibitemOpen
  \bibfield  {author} {\bibinfo {author} {\bibfnamefont {P.~M.}\ \bibnamefont
  {Lo}},\ }\href {\doibase 10.1103/PhysRevC.97.035210} {\bibfield  {journal}
  {\bibinfo  {journal} {Phys. Rev.}\ }\textbf {\bibinfo {volume} {C97}},\
  \bibinfo {pages} {035210} (\bibinfo {year} {2018})}\BibitemShut {NoStop}%
\bibitem [{\citenamefont {Huovinen}\ \emph {et~al.}(2017)\citenamefont
  {Huovinen}, \citenamefont {Lo}, \citenamefont {Marczenko}, \citenamefont
  {Morita}, \citenamefont {Redlich},\ and\ \citenamefont
  {Sasaki}}]{Huovinen:2016xxq}%
  \BibitemOpen
  \bibfield  {author} {\bibinfo {author} {\bibfnamefont {P.}~\bibnamefont
  {Huovinen}}, \bibinfo {author} {\bibfnamefont {P.~M.}\ \bibnamefont {Lo}},
  \bibinfo {author} {\bibfnamefont {M.}~\bibnamefont {Marczenko}}, \bibinfo
  {author} {\bibfnamefont {K.}~\bibnamefont {Morita}}, \bibinfo {author}
  {\bibfnamefont {K.}~\bibnamefont {Redlich}}, \ and\ \bibinfo {author}
  {\bibfnamefont {C.}~\bibnamefont {Sasaki}},\ }\href {\doibase
  10.1016/j.physletb.2017.03.060} {\bibfield  {journal} {\bibinfo  {journal}
  {Phys. Lett.}\ }\textbf {\bibinfo {volume} {B769}},\ \bibinfo {pages} {509}
  (\bibinfo {year} {2017})}\BibitemShut {NoStop}%
\bibitem [{\citenamefont {Vovchenko}\ \emph {et~al.}(2018)\citenamefont
  {Vovchenko}, \citenamefont {Gorenstein},\ and\ \citenamefont
  {Stoecker}}]{Vovchenko:2018fmh}%
  \BibitemOpen
  \bibfield  {author} {\bibinfo {author} {\bibfnamefont {V.}~\bibnamefont
  {Vovchenko}}, \bibinfo {author} {\bibfnamefont {M.~I.}\ \bibnamefont
  {Gorenstein}}, \ and\ \bibinfo {author} {\bibfnamefont {H.}~\bibnamefont
  {Stoecker}},\ }\href@noop {} {\  (\bibinfo {year} {2018})},\ \Eprint
  {http://arxiv.org/abs/1807.02079} {arXiv:1807.02079 [nucl-th]} \BibitemShut
  {NoStop}%
\bibitem [{\citenamefont {Andronic}\ \emph {et~al.}(2013)\citenamefont
  {Andronic}, \citenamefont {Braun-Munzinger}, \citenamefont {Redlich},\ and\
  \citenamefont {Stachel}}]{Andronic:2012dm}%
  \BibitemOpen
  \bibfield  {author} {\bibinfo {author} {\bibfnamefont {A.}~\bibnamefont
  {Andronic}}, \bibinfo {author} {\bibfnamefont {P.}~\bibnamefont
  {Braun-Munzinger}}, \bibinfo {author} {\bibfnamefont {K.}~\bibnamefont
  {Redlich}}, \ and\ \bibinfo {author} {\bibfnamefont {J.}~\bibnamefont
  {Stachel}},\ }\href {\doibase 10.1016/j.nuclphysa.2013.02.070} {\bibfield
  {journal} {\bibinfo  {journal} {Nucl. Phys.}\ }\textbf {\bibinfo {volume}
  {A904-905}},\ \bibinfo {pages} {535c} (\bibinfo {year} {2013})}\BibitemShut
  {NoStop}%
\bibitem [{\citenamefont {Becattini}\ and\ \citenamefont
  {Cleymans}(2007)}]{Becattini:2007qr}%
  \BibitemOpen
  \bibfield  {author} {\bibinfo {author} {\bibfnamefont {F.}~\bibnamefont
  {Becattini}}\ and\ \bibinfo {author} {\bibfnamefont {J.}~\bibnamefont
  {Cleymans}},\ }\href {\doibase 10.1088/0954-3899/34/8/S135} {\bibfield
  {journal} {\bibinfo  {journal} {J. Phys.}\ }\textbf {\bibinfo {volume}
  {G34}},\ \bibinfo {pages} {S959} (\bibinfo {year} {2007})}\BibitemShut
  {NoStop}%
\bibitem [{\citenamefont {Rakotozafindrabe}\ \emph {et~al.}(2013)\citenamefont
  {Rakotozafindrabe} \emph {et~al.}}]{Rakotozafindrabe:2012ei}%
  \BibitemOpen
  \bibfield  {author} {\bibinfo {author} {\bibfnamefont {A.}~\bibnamefont
  {Rakotozafindrabe}} \emph {et~al.},\ }\href {\doibase
  10.1016/j.nuclphysa.2013.02.174} {\bibfield  {journal} {\bibinfo  {journal}
  {Nucl. Phys.}\ }\textbf {\bibinfo {volume} {A904-905}},\ \bibinfo {pages}
  {957c} (\bibinfo {year} {2013})}\BibitemShut {NoStop}%
\bibitem [{\citenamefont {Brodsky}\ \emph {et~al.}(2013)\citenamefont
  {Brodsky}, \citenamefont {Fleuret}, \citenamefont {Hadjidakis},\ and\
  \citenamefont {Lansberg}}]{Brodsky:2012vg}%
  \BibitemOpen
  \bibfield  {author} {\bibinfo {author} {\bibfnamefont {S.~J.}\ \bibnamefont
  {Brodsky}}, \bibinfo {author} {\bibfnamefont {F.}~\bibnamefont {Fleuret}},
  \bibinfo {author} {\bibfnamefont {C.}~\bibnamefont {Hadjidakis}}, \ and\
  \bibinfo {author} {\bibfnamefont {J.~P.}\ \bibnamefont {Lansberg}},\ }\href
  {\doibase 10.1016/j.physrep.2012.10.001} {\bibfield  {journal} {\bibinfo
  {journal} {Phys. Rept.}\ }\textbf {\bibinfo {volume} {522}},\ \bibinfo
  {pages} {239} (\bibinfo {year} {2013})}\BibitemShut {NoStop}%
\bibitem [{\citenamefont {Karpenko}(2018)}]{Karpenko:2018xam}%
  \BibitemOpen
  \bibfield  {author} {\bibinfo {author} {\bibfnamefont {I.}~\bibnamefont
  {Karpenko}},\ }\href@noop {} {\  (\bibinfo {year} {2018})},\ \Eprint
  {http://arxiv.org/abs/1805.11998} {arXiv:1805.11998 [nucl-th]} \BibitemShut
  {NoStop}%
\bibitem [{\citenamefont {Begun}\ \emph {et~al.}(2018)\citenamefont {Begun},
  \citenamefont {Kikoła}, \citenamefont {Vovchenko},\ and\ \citenamefont
  {Wielanek}}]{Begun:2018efg}%
  \BibitemOpen
  \bibfield  {author} {\bibinfo {author} {\bibfnamefont {V.}~\bibnamefont
  {Begun}}, \bibinfo {author} {\bibfnamefont {D.}~\bibnamefont {Kikoła}},
  \bibinfo {author} {\bibfnamefont {V.}~\bibnamefont {Vovchenko}}, \ and\
  \bibinfo {author} {\bibfnamefont {D.}~\bibnamefont {Wielanek}},\ }\href@noop
  {} {\  (\bibinfo {year} {2018})},\ \Eprint {http://arxiv.org/abs/1806.01303}
  {arXiv:1806.01303 [nucl-th]} \BibitemShut {NoStop}%
\bibitem [{\citenamefont {Adriani}\ \emph {et~al.}(2015)\citenamefont {Adriani}
  \emph {et~al.}}]{Adriani:2015nwa}%
  \BibitemOpen
  \bibfield  {author} {\bibinfo {author} {\bibfnamefont {O.}~\bibnamefont
  {Adriani}} \emph {et~al.} (\bibinfo {collaboration} {LHCf}),\ }\href
  {\doibase 10.1016/j.physletb.2015.09.041} {\bibfield  {journal} {\bibinfo
  {journal} {Phys. Lett.}\ }\textbf {\bibinfo {volume} {B750}},\ \bibinfo
  {pages} {360} (\bibinfo {year} {2015})},\ \Eprint
  {http://arxiv.org/abs/1503.03505} {arXiv:1503.03505 [hep-ex]} \BibitemShut
  {NoStop}%
\bibitem [{\citenamefont {Itow}\ \emph {et~al.}(2014)\citenamefont {Itow} \emph
  {et~al.}}]{Itow:2014oea}%
  \BibitemOpen
  \bibfield  {author} {\bibinfo {author} {\bibfnamefont {Y.}~\bibnamefont
  {Itow}} \emph {et~al.},\ }\href@noop {} {\  (\bibinfo {year} {2014})},\
  \Eprint {http://arxiv.org/abs/1409.4860} {arXiv:1409.4860 [physics.ins-det]}
  \BibitemShut {NoStop}%
\bibitem [{\citenamefont {Brewer}\ \emph {et~al.}(2018)\citenamefont {Brewer},
  \citenamefont {Mukherjee}, \citenamefont {Rajagopal},\ and\ \citenamefont
  {Yin}}]{Brewer:2018abr}%
  \BibitemOpen
  \bibfield  {author} {\bibinfo {author} {\bibfnamefont {J.}~\bibnamefont
  {Brewer}}, \bibinfo {author} {\bibfnamefont {S.}~\bibnamefont {Mukherjee}},
  \bibinfo {author} {\bibfnamefont {K.}~\bibnamefont {Rajagopal}}, \ and\
  \bibinfo {author} {\bibfnamefont {Y.}~\bibnamefont {Yin}},\ }\href@noop {} {\
   (\bibinfo {year} {2018})},\ \Eprint {http://arxiv.org/abs/1804.10215}
  {arXiv:1804.10215 [hep-ph]} \BibitemShut {NoStop}%
\end{thebibliography}%

\end{document}